\documentstyle[prl,aps,epsfig]{revtex}
\def\bi{\bibitem}
\def\prl{Phys. Rev. Lett.}\def\np{Nucl. Phys.}\def\pr{Phys. Rev.}
\def\pl{Phys. Lett.}
\def\ba{\begin{eqnarray}}
\def\ea{\end{eqnarray}}
\def\M{{\cal M}}\def\la{\langle}\def\ra{\rangle}
\def\no{\nonumber\\}

\begin{document}
\twocolumn[%
\hsize\textwidth\columnwidth\hsize\csname@twocolumnfalse\endcsname
\hfill{\normalsize\vbox{\hbox{KIAS-P01048}  }}

\vskip 0.5cm
\title{%
\bf Vector Manifestation and Fate of Vector Mesons in Dense
Matter}
\author{{\bf Masayasu Harada$^{(a,b)}$}, {\bf Youngman
Kim$^{(c)}$}
and {\bf Mannque Rho$^{(d,b)}$}}
\address{(a) Department of Physics, Nagoya University,
Nagoya, 464-8602, Japan\\ (b) School of Physics, Korea Institute
for Advanced Study, Seoul 130-012, Korea\\ (c) School of Physics,
Seoul National University, Seoul 151-742, Korea\\ (d) Service de
Physique Th\'eorique, CEA Saclay, 91191 Gif-sur-Yvette, France}

 \maketitle
\vskip 0.2cm

\centerline{(\today)}

\begin{abstract}
We describe in-medium properties of hadrons in dense matter near
chiral restoration using a Wilsonian matching to QCD of an
effective field theory with hidden local symmetry at the chiral
cutoff $\Lambda$. We find that chiral symmetry is restored in
vector manifestation \`a la Harada and Yamawaki at a critical
matter density $n_c$. We express the critical density in terms of
QCD correlators in dense matter at the matching scale. In a manner
completely analogous to what happens at the critical $N_f^c$ and
at the critical temperature $T_c$, the vector meson mass is found
to vanish (in the chiral limit) at chiral restoration. This result
provides a support for Brown-Rho scaling predicted a decade ago.
\end{abstract}
\vskip1pc]

\vspace{-0.2cm}

Following recent developments on hidden local
symmetry~\cite{HY:VM} and color-flavor locking~\cite{wetterich} in
the hadronic sector, Brown and Rho proposed
\cite{BR:Berkeley,BR:PR2001} that the vector manifestation (VM)
scenario of Harada and Yamawaki~\cite{HY:VM} for the realization
of chiral symmetry in strongly interacting systems which was shown
to be valid for large number of flavors $N_f$ should be also
applicable to high density (or high temperature) hadronic matter
relevant to the interior of compact stars (or relativistic
heavy-ion process) and that as a consequence, the scaling behavior
of vector mesons in medium proposed by Brown and
Rho~\cite{BRscaling} near chiral restoration critical density
$n_c$ (or temperature $T_c$) follows from VM. That the vector
meson mass vanishes in the chiral limit at the critical
temperature $T_c$ in accordance with the VM mode was recently
shown to hold by Harada and Sasaki~\cite{HS:T}. In this paper, we
supply the arguments to suggest that the same phenomenon occurs in
density, namely that at $n=n_c$, the vector meson mass vanishes in
the chiral limit.

We begin by giving a brief summary of the key arguments as to how
VM figures in the properties of hadrons in medium.

To study how hadrons behave in dense (hot) medium starting from
normal conditions, one resorts to effective field theories with
Lagrangians that have the assumed symmetry properties of QCD. Such
Lagrangians are constructed so as to describe low-energy
interactions of hadrons in medium-free vacuum. As one increases
the density (temperature), that is, as the scale is changed, the
flow of the given theory is not unique even though the symmetries
remain unchanged. As shown by Harada and Yamawaki~\cite{HY:fate},
the effective field theory with hidden local symmetry
(HLS)~\cite{BKUYY,BKY:88} can flow to two or more different fixed
points
depending upon how the $parameters$ of the Lagrangian are dialed.
It turns out that if the $bare$ parameters of the Lagrangian are
matched \`a la Wilson to QCD at the chiral scale $\Lambda_\chi\sim
4\pi f_\pi$  (where $f_\pi$ is the pion decay constant) above the
vector meson mass -- based on the systematic chiral perturbation with
HLS~\cite{HY:PLB,Tanabashi,HY:matching,HY:PR}, then the flow
comes out to be unique with the fixed point~\cite{HY:fate}.
This implies that the
different flows typically present in all effective field theories,
even if consistent with the symmetries of QCD, may not correctly
represent QCD dynamics unless the $bare$ parameters of the
effective Lagrangian are matched at an appropriate scale, say,
$\Lambda_\chi$, with QCD. Most remarkably though, when the HLS
theory is matched with QCD, Wilsonian renormalization group
equations (RGEs) show that the mass $parameter$ $M_\rho$ of the
vector meson and the hidden gauge coupling $parameter$ $g$ do flow
to zero together with the pion decay constant $f_\pi$ going to
zero at the chiral restoration point realizing what is referred to
as ``vector manifestation (VM)'' and consequently the vector meson
pole mass which is given in terms of the parameters $M_\rho$ and
$g$ vanishes at the critical point with the multiplet of vector
mesons decoupling. This has been shown to be what happens at
$N_f=N_f^c\sim 5$ for $T=0$~\cite{HY:VM} and $T=T_c\sim 250$ MeV
for $N_f=3$~\cite{HS:T}.

To set up the arguments for the density problem, we consider a
system of hadrons in the background of a filled Fermi sea. For the
moment, we consider the Fermi sea as merely a $background$,
side-stepping the question of how the Fermi sea is formed from a
theory defined in a matter-free vacuum. Imagine that mesons -- the
pion and the $\rho$ meson -- are introduced in HLS
theory~\cite{BKUYY,BKY:88} with a cutoff set at the scale, say,
$\Lambda_\chi$. Since we are dealing with dense fermionic matter,
we will need to introduce the degrees of freedom associated with
baryons or alternatively constituent quarks (or quasiquarks). At
low density, say, $n < \tilde{n}$, with $\tilde{n}$ being some
density greater than $n_0$, the precise value of which cannot be
pinned down at present, we may choose to put the cutoff
$\Lambda_0$ below the nucleon mass, $m_N\sim 1$ GeV, but above the
$\rho$ mass $m_\rho=770$ MeV and integrate out all the baryons. In
this case, the $bare$ parameters of the HLS Lagrangian will depend
upon the density $n$ (or equivalently Fermi momentum $P_F$) since
the baryons that are integrated out carry information about the
baryon density through their interactions in the full theory with
the baryons within the Fermi sea. Once the baryons are integrated
out, we will then be left with the standard HLS Lagrangian theory
with the NG and gauge boson fields only {\it except that the bare
parameters of the effective Lagrangian will be density-dependent}.
It should be noticed that {\it the cutoff can also be density
dependent}. However, in general, the density-dependence of the
cutoff is not related to those of the bare parameters by the RGEs.
For $T> 0$ and $n=0$ this difference appears from the
``intrinsic'' temperature dependence introduced in
Ref.~\cite{HS:T} which was essential for the VM to occur at the
chiral restoration point.

As density increases beyond $\tilde{n}$, the fermions will however
start figuring explicitly, that is, the fermion field will be
present below the cutoff $\tilde{\Lambda} (n>\tilde{n})$. The
reason is that as density approaches the chiral restoration point,
the constituent-quark (called quasiquark) picture -- which seems
to be viable even in matter-free space~\cite{swenson} -- becomes
more appropriate~\cite{BR:PR2001} and the quasiquark mass drops
rapidly, ultimately vanishing (in the chiral limit) at the
critical point. This picture has been advocated by several authors
in a related context~\cite{riska-brown}.

We now describe in some detail how the above scenario takes place.
As a simple albeit unrealistic case in dense matter,
consider the fermionic
degrees of freedom to be baryons with a mass scale above the
cutoff for all densities up to the chiral restoration density. In
this case we can integrate out the baryons and take, as in
\cite{HY:VM,HS:T,HY:fate}, the standard HLS model based on the
$G_{\rm global} \times H_{\rm local}$ symmetry, where $G =
\mbox{SU($N_f$)}_{\rm L} \times \mbox{SU($N_f$)}_{\rm R}$  is the
global chiral symmetry and $H = \mbox{SU($N_f$)}_{\rm V}$ is the
HLS. When the kinetic term of gauge bosons of $H_{\rm local}$ is
ignored, the HLS model is reduced to the nonlinear sigma model
based on $G/H$, with $G_{\rm global}\times H_{\rm local}$ broken
down to the diagonal sum which is the flavor symmetry $H$ of
$G/H$. In the HLS model the basic quantities are the gauge bosons
$\rho_\mu = \rho_\mu^a T_a$ of the HLS and two SU($N_f$)-matrix
valued variables $\xi_{\rm L}$ and $\xi_{\rm R}$. They are
parameterized as $ \xi_{\rm L,R} = e^{i\sigma/F_\sigma} e^{\mp
i\pi/F_\pi} $, where $\pi = \pi^a T_a$ denote the pseudoscalar
Nambu-Goldstone
(NG) bosons associated with the spontaneous breaking of $G$ and
$\sigma = \sigma^a T_a$ the NG bosons absorbed into the HLS gauge
bosons $\rho_\mu$ which are identified with the vector mesons.
$F_\pi$ and $F_\sigma$ are relevant decay
constants, and the parameter $a$ is defined as $a \equiv
F_\sigma^2/F_\pi^2$.
$\xi_{\rm L}$ and $\xi_{\rm R}$ transform as $\xi_{\rm L,R}(x)
\rightarrow h(x) \xi_{\rm L,R}(x) g^{\dag}_{\rm L,R}$, where $h(x)
\in H_{\rm local}$ and $g_{\rm L,R} \in G_{\rm global}$. The
covariant derivatives of $\xi_{\rm L,R}$ are defined by $ D_\mu
\xi_{\rm L} =
\partial_\mu \xi_{\rm L} - i g \rho_\mu \xi_{\rm L}
+ i \xi_{\rm L} {\cal L}_\mu $, and similarly with replacement
${\rm L} \leftrightarrow {\rm R}$, ${\cal L}_\mu \leftrightarrow
{\cal R}_\mu$, where $g$ is the HLS gauge coupling, and ${\cal
L}_\mu$ and ${\cal R}_\mu$ denote the external gauge fields
gauging the $G_{\rm global}$ symmetry. The HLS Lagrangian is given
by~\cite{BKUYY,BKY:88}
\begin{equation}
{\cal L} = F_\pi^2 \, \mbox{tr} \left[ \hat{\alpha}_{\perp\mu}
\hat{\alpha}_{\perp}^\mu \right] + F_\sigma^2 \, \mbox{tr} \left[
  \hat{\alpha}_{\parallel\mu} \hat{\alpha}_{\parallel}^\mu
\right] + {\cal L}_{\rm kin}(\rho_\mu) \ ,
\label{Lagrangian}
\end{equation}
where ${\cal L}_{\rm kin}(\rho_\mu)$ denotes the kinetic term of
$\rho_\mu$ and
\begin{eqnarray}
&&
\hat{\alpha}_{\stackrel{\perp}{\scriptscriptstyle\parallel}}^\mu =
\left(
  D_\mu \xi_{\rm R} \cdot \xi_{\rm R}^\dag \mp
  D_\mu \xi_{\rm L} \cdot \xi_{\rm L}^\dag
\right) / (2i) \ .
\end{eqnarray}

As stated above, the three parameters of the Lagrangian $F_\pi$,
$F_\sigma$ (or $a$) and $g$ will depend on density. Since Lorentz
invariance is broken, distinction has to be made between the
temporal and spatial components of the constants in
Eq.~(\ref{Lagrangian}). We will ignore the difference for the
moment. This will be justified below and, in more detail, in
Appendix A. For the moment continuing with Eq.~(\ref{Lagrangian}),
we need to match Eq.~(\ref{Lagrangian}) with QCD to define the
$bare$ Lagrangian for the effective theory. To determine the
$bare$ parameters, we set the matching scale at
$\Lambda\approx\Lambda_\chi$ below which only the HLS degrees of
freedom are present and extend the Wilsonian
matching~\cite{HY:matching}, which was originally proposed for
$T=n=0$ in Ref.~\cite{HY:matching} and extended to non-zero
temperature in Ref.~\cite{HS:T}, to non-zero density. We match the
axial-vector and vector-current correlators in the HLS with those
derived in the OPE for QCD at non-zero density. The correlators in
the HLS around the matching scale $\M=\Lambda$ (where $\M$ is the
renormalization scale~\footnote{We reserve $\mu$ for chemical
potential.}) are well described by the same forms as those at
$T=n=0$~\cite{HY:matching} with the bare parameters having the
``intrinsic'' density dependence~\footnote{Note that at the level
of the $bare$ Lagrangian, there is no vector-axial-vector mixing
discussed for hot matter by Dey, Eletsky and Ioffe~\cite{mixing}.
At the matching scale, there are no loop corrections. Mixing occurs
through hadronic loops when decimation is made. }:
\begin{eqnarray}
\Pi_A^{\rm(HLS)}(Q^2) &=& \frac{F_\pi^2(\Lambda;n)}{Q^2} - 2
z_2(\Lambda;n) \ ,
\nonumber\\
\Pi_V^{\rm(HLS)}(Q^2) &=& \frac{
  F_\sigma^2(\Lambda;n)
  \left[ 1 - 2 g^2(\Lambda;n) z_3(\Lambda;n) \right]
}{
  M_\rho^2(\Lambda;n) + Q^2
}
\nonumber\\
&& - 2 z_1(\Lambda;n) \ , \label{Pi A V HLS}
\end{eqnarray}
where $M_\rho^2(\Lambda;n) \equiv g^2(\Lambda;n)
F_\sigma^2(\Lambda;n)$ is the bare $\rho$ mass, and
$z_{1,2,3}(\Lambda;n)$ are the bare coefficient parameters of the
relevant ${\cal O}(p^4)$ terms~\cite{Tanabashi,HY:matching}, all
at $\M=\Lambda$. Since the Lorentz non-invariant terms in the
current correlators by the OPE are suppressed by some powers of
$n/\Lambda^3$ (see, e.g. Ref.~\cite{Hatsuda-Lee}), we can ignore
them from both the hadronic and QCD sectors. (See Appendix A for
the justification for the hadronic sector).
 Matching the above correlators with those by
the OPE in the same way as done for $T=n=0$~\cite{HY:matching}, we
determine the bare parameters that include what we shall call
``intrinsic'' density dependence, which are then converted into
those of the on-shell parameters through the Wilsonian
RGE's~\cite{HY:VM,HY:matching}. As a result, the parameters
appearing in the hadronic density corrections have the intrinsic
density dependence.

Now, to study the chiral restoration in dense matter, we assume
that we can do in the fermion-less theory the Wilsonian matching
at the critical density $n_c$ for $N_f = 3$ assuming that $\langle
\bar{q} q \rangle$ approaches to $0$ (continuously)\footnote{%
We are assuming that the transition is not strongly first order.
The quasiquark degrees of freedom introduced later make sense only
within the same hypothesis. There is nothing at present that
invalidates our assumption, but if the transition were proven to
be strongly first order, some of the arguments used in this paper
might need qualifications. We note that, in the presence of the
current quark mass, the quark condensate is believed to decrease
rapidly but continuously around the ``phase transition''
point~\cite{BR:PRep}. } for $n \rightarrow n_c$. Then, the
axial-vector and vector-current correlators given by OPE in the
QCD sector approach each other, and will agree at $n_c$. Then
through the Wilsonian matching we require that the correlators in
Eq.~(\ref{Pi A V HLS}) agree with each other. As in the case of
large $N_f$~\cite{HY:VM} and in the case of $T\sim
T_c$~\cite{HS:T}, this agreement can be satisfied also in dense
matter if the following conditions are met:
\begin{eqnarray}
&& g(\Lambda;n) \mathop{\longrightarrow}_{n \rightarrow n_c} 0 \ ,
\qquad a(\Lambda;n) \mathop{\longrightarrow}_{n \rightarrow n_c} 1
\ ,
\nonumber\\
&& z_1(\Lambda;n) - z_2(\Lambda;n) \mathop{\longrightarrow}_{n
\rightarrow n_c} 0 \ .
\label{g a z12:VMT}
\end{eqnarray}
We show in Appendix A that these conditions remain valid -- with a
suitable in-medium extension  -- when the breaking of Lorentz
symmetry in medium is taken into account in the bare Lagrangian.

Next we need to consider how these parameters flow as the scale
parameter is varied. The flows are obtained by solving the RGEs
for the parameters. The RGEs for the parameters of the HLS theory
as the scale ${\cal M}$ is varied were derived in
Refs.~\cite{HY:conformal,HY:VM,HY:fate} with the effect of
quadratic divergences included. These equations describe the flow
of the parameters for dense system for a {\it fixed} chemical
potential $\mu$ (or density $n$)\footnote{We are using density $n$
and chemical potential $\mu$ interchangeably. In the case of
nearly massless quasiquarks near chiral restoration, $\mu\approx
P_F$.}. They show that $a=1$, $g=0$ and $X=1$ with $X$ defined by
 \ba
X\equiv \frac{N_f}{2(4\pi)^2}\frac{\M^2}{F_\pi^2(\M)}
  \label{def:X}
 \ea
are fixed points. Thus at $\mu=\mu_c$, given the bare parameters
(\ref{g a z12:VMT}) at the matching scale $\Lambda$, both $g$ and
$a$ flow to the fixed point. The RGEs given in
Refs.~\cite{HY:conformal,HY:VM,HY:fate} then imply that at
$\mu=\mu_c$, $g=0$ and $a=1$ remain unchanged as $\M$ is varied.
Now what about $F_\pi (\M)$ which cannot be fixed by requiring
only the agreement between the vector and axial-vector current
correlators? As we will discuss in more detail later, in the
absence of the hadronic dense-loop corrections $F_\pi
(\M=0;\mu_c)=0$ is obtained from the fact that $X=1$ is a fixed
point and corresponds to the pion decay constant $f_\pi
(\mu_c)=0$. {\it Thus the chiral transition in high density will
coincide with the VM}.

As stated, as density is raised --  and in particular near the
critical density on which we will focus, we expect the fermionic
degrees of freedom to figure explicitly below the cutoff at which
the Wilsonian matching is effected. In principle, to account for
the fermionic degrees of freedom below the scale $\M=\Lambda
(\mu)$ for a given $\mu > \tilde{\mu}$, we may introduce either
light baryons with a running mass that drops with density or, more
appropriately, constituent quarks with masses scaling with density
as suggested by Riska and Brown~\cite{riska-brown}. We adopt the
latter in this paper.

We introduce the quasiquark field $\psi$ below the scale $\Lambda
(\mu)$ for $\mu\geq \tilde{\mu}$ into the Lagrangian. A chiral
Lagrangian for $\pi$ with the constituent quark (quasiquark) was
given in Ref.~\cite{MG}. In Ref.~\cite{BKY:88} the quasiquark
field, say $\psi$, is introduced into the HLS Lagrangian in such a
way that it transforms homogeneously under the HLS: $\psi
\rightarrow h(x)\cdot \psi$ where $h(x) \in H_{\rm local}$. Here
we extend the Lagrangian of Ref.~\cite{BKY:88} to a general one
with which we can perform a systematic derivative expansion. Since
we are considering the model near the chiral phase transition
point where the quasiquark mass is expected to become small, we
assign ${\cal O}(p)$ to the constituent-quark (quasiquark) mass
$m_q$. Furthermore, we assign ${\cal O}(p)$ to the chemical
potential $\mu$ or the Fermi momentum $P_F$, as we consider that
the cutoff is larger than $\mu$ even near the phase transition
point. Using this counting scheme we can make the systematic
expansion in the HLS with the quasiquark included. We should note
that this counting scheme is different from the one in the model
for $\pi$ and baryons given in Ref.~\cite{MOR:01} where the baryon
mass is counted as ${\cal O}(1)$. The leading order Lagrangian
including one quasiquark field and one anti-quasiquark field is
counted as ${\cal O}(p)$ and given by
 \begin{eqnarray}
 \delta {\cal L}_{Q(1)} &=& \bar \psi(x)( iD_\mu \gamma^\mu
       - \mu \gamma^0 -m_q )\psi(x)\nonumber\\
 &&+ \bar \psi(x) \left(
  \kappa\gamma^\mu \hat{\alpha}_{\parallel \mu}(x )
+ \lambda\gamma_5\gamma^\mu \hat{\alpha}_{\perp\mu}(x) \right)
        \psi(x) \label{lagbaryon}
 \end{eqnarray}
where $D_\mu\psi=(\partial_\mu -ig\rho_\mu)\psi$ and $\kappa$ and
$\lambda$ are constants to be specified later. At one-loop level
the Lagrangian (\ref{lagbaryon}) generates the ${\cal O}(p^4)$
contributions including hadronic dense-loop effects as well as
divergent effects. The divergent contributions are renormalized by
the parameters, and thus the RGEs for three leading order
parameters $F_\pi$, $a$ and $g$ (and parameters of ${\cal O}(p^4)$
Lagrangian) are modified from those without quasiquark field. In
addition, we need to consider the renormalization group flow for
the quasiquark mass $m_q$~\footnote{The constants $\kappa$ and
$\lambda$ will also run such that at $\mu=\mu_c$,
$\kappa=\lambda=1$ while at $\mu < \mu_c$, $\kappa\neq \lambda$.
The running will be small near $n_c$, so we will ignore their
running here.}. Calculating one-loop contributions for RGEs in
$\M$ for a given $\mu$, we find
 \ba
\M \frac{dF_\pi^2}{d\M} &=& C[3a^2g^2F_\pi^2
+2(2-a)\M^2] -\frac{m_q^2}{2\pi^2}\lambda^2 N_c\nonumber\\
\M \frac{da}{d\M} &=&-C
(a-1)[3a(1+a)g^2-(3a-1)\frac{\M^2}{F_\pi^2}]\nonumber\\
&& +a\frac{\lambda^2}{2\pi^2}\frac{m_q^2}{F_\pi^2}N_c\nonumber\\
\M\frac{d g^2}{d \M}&=& -C\frac{87-a^2}{6}g^4
+\frac{N_c}{6\pi^2}g^4 (1-\kappa)^2\nonumber\\
\M\frac{dm_q}{d\M}&=& -\frac{m_q}{8\pi^2}[ (C_\pi-C_\sigma)\M^2
-m_q^2 (C_\pi-C_\sigma)\nonumber\\
&& +M_\rho^2C_\sigma -4C_\rho ]\label{rgewbaryons}
 \ea
where $C = N_f/\left[2(4\pi)^2\right]$ and
 \ba
C_\pi&\equiv
&(\frac{\lambda}{F_\pi})^2
\frac{N_f^2 - 1}{2N_f}
\nonumber\\
C_\sigma&\equiv
&(\frac{\kappa}{F_\sigma})^2
\frac{N_f^2 - 1}{2N_f}
\nonumber\\
C_\rho&\equiv & g^2 (1-\kappa)^2
\frac{N_f^2 - 1}{2N_f}
\nonumber
 \ea

For $\mu >\tilde{\mu}$ at which the quasiquarks enter, the cutoff
will be different from that without. However the matching
conditions (\ref{g a z12:VMT}) will remain the same. Now
eq.~(\ref{rgewbaryons}) shows that $(g\,,\,a)=(0\,,\,1)$ is a
fixed point only when $m_q=0$. Since $m_q=0$ itself is a fixed
point of the RGE for $m_q$, $(g\,,\,a\,,\,m_q) = (0\,,\,1\,,\,0)$
is a fixed point of the coupled RGEs for $g$, $a$ and $m_q$.
Furthermore and most importantly, {\it $X=1$ becomes the fixed
point of the RGE for} $X$~\cite{HY:fate}. This means that at the
fixed point, $F_\pi (0)=0$ [see Eq.~(\ref{def:X})]. What does this
mean in dense matter? To see what this means, we note that for
$T=\mu=0$, this $F_\pi (0)=0$ condition is satisfied for a given
number of flavors $N_f^{\rm cr}\sim 5$ through the Wilsonian
matching~\cite{HY:VM}. For $N_f=3$, $\mu=0$ and $T\neq 0$, this
condition is never satisfied due to thermal hadronic
corrections~\cite{HS:T}. Remarkably, as we show in Appendix B, for
$N_f=3$, $T=0$ and $\mu=\mu_c$, it turns out that dense hadronic
corrections vanish up to ${\cal O}(p^6)$ corrections. Therefore
the fixed point $X=1$ (i.e., $F_\pi (0)=0$) does indeed signal
chiral restoration at the critical density.

Let us here
focus on what
happens to hadrons at and very near the critical point $\mu_c$.
This problem can be easily addressed with the machinery developed
above. To do this we define, following \cite{HS:T}, the
``on-shell" quantities
\begin{eqnarray}
&&
  F_\pi = F_\pi(\M=0;\mu) \ ,
\nonumber\\
&&
  g = g\mbox{\boldmath$\bigl($}
  \M = M_\rho(\mu);\mu
  \mbox{\boldmath$\bigr)$}
\ , \quad
  a = a\mbox{\boldmath$\bigl($}
  \M = M_\rho(\mu);\mu
  \mbox{\boldmath$\bigr)$}
\ , \label{on-shell para mu}
\end{eqnarray}
where $M_\rho$ is determined from the ``on-shell condition":
\begin{eqnarray}
&&
  M_\rho^2 = M_\rho^2(\mu) =
  a\mbox{\boldmath$\bigl($}
  \M = M_\rho(\mu);\mu
  \mbox{\boldmath$\bigr)$}
\nonumber\\
&& \quad
  \times
  g^2\mbox{\boldmath$\bigl($}
  \M = M_\rho(\mu);\mu
  \mbox{\boldmath$\bigr)$}
  F_\pi^2\mbox{\boldmath$\bigl($}
  \M = M_\rho(\mu);\mu
  \mbox{\boldmath$\bigr)$}
\ .
\end{eqnarray}
Then, the parameter $M_\rho$ in this paper is renormalized in such
a way that it becomes the pole mass at $\mu=0$.

We first look at the ``on-shell'' pion decay constant $f_\pi$. At
$\mu=\mu_c$, it is given by
 \ba
f_\pi (\mu_c)\equiv f_\pi (\M=0;\mu_c)=F_\pi (0;\mu_c) + \Delta
(\mu_c)
 \ea
where $\Delta$ is dense hadronic contribution arising from fermion
loops involving (\ref{lagbaryon}). As we shall show explicitly in
Appendix B, up to ${\cal O}(p^6)$ in the power counting,
$\Delta(\mu_c)=0$ at the fixed point $(g,a,m_q)=(0,1,0)$. Thus
 \ba
f_\pi (\mu_c)=F_\pi (0;\mu_c)=0.
 \ea
Since
 \ba
F_\pi^2 (0;\mu_c)=F_\pi^2
(\Lambda;\mu_c)-\frac{N_f}{2(4\pi)^2}\Lambda^2,
 \ea
and at the matching scale $\Lambda$, $F_\pi^2 (\Lambda;\mu_c)$ is
given by a QCD correlator at $\mu=\mu_c$,  $\mu_c$ can be computed
from
 \ba
  F_\pi^2 (\Lambda;\mu_c)=\frac{N_f}{2(4\pi)^2}\Lambda^2 \ .
 \ea
Note that in free space, this is the equation that determines
$N_f^c\sim 5$~\cite{HY:VM}. In order for this equation to have a
solution at the critical density, it is necessary that $F_\pi^2
(\Lambda;\mu_c)/F_\pi^2 (\Lambda;0) \sim 3/5$. We do not have at
present a reliable estimate of the density dependence of the QCD
correlator to verify this condition but the decrease of $F_\pi$ of
this order in medium looks quite reasonable.

Next we compute the $\rho$ pole mass near $\mu_c$. The details of
the calculation are given in Appendix C. Here we just quote the
result. With the inclusion of the fermionic dense loop terms, the
pole mass, for $M_\rho, m_q \ll P_F$, is  of the form
 \ba
 m_\rho^2(\mu) &=& M_\rho^2 (\mu) + g^2
  \, G(\mu)\ ,\label{mrho at T 2}\\
 G(\mu)&=&\frac{\mu^2}{2\pi^2} [\frac 13 (1-\kappa)^2 +N_c
 (N_fc_{V1}+c_{V2})]\ .
  \ea
At $\mu=\mu_c$, we have $g=0$ and $a=1$ so that $M_\rho (\mu)=0$
and since $G(\mu_c)$ is non-singular, $m_\rho=0$. Thus the fate of
the $\rho$ meson at the critical density is the same as that at
the critical temperature. This is our main result. It is
noted~\cite{HY:VM} that although the conditions for $g(\Lambda;n)$
and $a(\Lambda;n)$ in Eq.~(\ref{g a z12:VMT}) coincided with the
Georgi's vector limit~\cite{Georgi}~\footnote{It was suggested in
\cite{BR94} that the Georgi vector limit was relevant to chiral
restoration. Here we note that the chiral transition involves {\it
both} the vector limit and the vanishing of the pion decay
constant. A non-zero pion decay constant with the vector limit is
not consistent with low-energy theorems.}, the VM should be
distinguished from Georgi's vector realization~\cite{Georgi}.

So far we have focused on the critical density at which the
Wilsonian matching clearly determines $g=0$ and $a=1$ without
knowing much about the details of the current correlators. Here we
consider how the parameters flow as function of chemical potential
$\mu$. In the low density region, we expect that the ``intrinsic''
density dependence of the bare parameters is small. If we ignore
the intrinsic density effect, we may then resort to
Morley-Kislinger (MS) theorem 2~\cite{morley-kislinger} (sketched
and referred for definiteness to as ``MK theorem" in Appendix D)
which states that given an RGE in terms of $\M$, one can simply
trade in $\mu$ for $\M$ for dimensionless quantities and for
dimensionful quantities with suitable calculable additional terms.
The results are
 \ba
\mu \frac{dF_\pi^2}{d\mu} &=& -2F_\pi^2 + C[3a^2g^2F_\pi^2
+2(2-a)\M^2] -\frac{m^2}{2\pi^2}\lambda^2 N_c\nonumber\\
\mu \frac{da}{d\mu} &=&-C
(a-1)[3a(1+a)g^2-(3a-1)\frac{\mu^2}{F_\pi^2}]\nonumber\\
&& +a\frac{\lambda^2}{2\pi^2}\frac{m^2}{F_\pi^2}N_c\nonumber\\
\mu\frac{d g^2}{d \mu}&=& -C\frac{87-a^2}{6}g^4
+\frac{N_c}{6\pi^2}g^4 (1-\kappa)^2\nonumber\\
\mu\frac{dm_q}{d\mu}&=& -m_q -\frac{m_q}{8\pi^2}[
(C_\pi-C_\sigma)\mu^2
-m_q^2 (C_\pi-C_\sigma)\nonumber\\
&& +M_\rho^2C_\sigma -4C_\rho ]
 \ ,
\label{RGE-mu}
\end{eqnarray}
where $F_\pi$, $a$, $g$, etc.~are understood as
$F_\pi({\cal M}=\mu;\mu)$,
$a({\cal M}=\mu;\mu)$,
$g({\cal M}=\mu;\mu)$, and so on.

It should be stressed that the MK theorem presumably applies in
the given form to ``fundamental theories" such as QED but not
without modifications to effective theories such as the one we are
considering. The principal reason is that there is a change of
relevant degrees of freedom from above $\Lambda$ where QCD
variables are relevant to below $\Lambda$ where hadronic variables
figure. Consequently we do not expect Eq.~(\ref{RGE-mu}) to apply
in the vicinity of $\mu_c$. Specifically, near the critical point,
the ``intrinsic'' density dependence of the bare theory will
become indispensable and the naive application of
Eq.~(\ref{RGE-mu}) should break down. One can see this clearly in
the following example: The condition $g(\M=\mu_c;\mu_c)=0$ that
follows from the matching condition (\ref{Pi A V HLS}), would
imply, when (\ref{RGE-mu}) is naively applied, that ${g}(\mu)=0$
for {\it all} $\mu$. This is obviously incorrect~\footnote{The
RGEs (\ref{RGE-mu}) were recently studied in \cite{kl99} with the
nucleons incorporated as explicit fermionic degrees of freedom.}.
Near the critical density the ``intrinsic'' density dependence
should be included in the RGE: Noting that Eq.~(\ref{RGE-mu}) is
for, e.g., $g({\cal M}=\mu;\mu)$, we can write down the RGE for
$g$ corrected by the ``intrinsic'' density dependence as
\begin{eqnarray}
&&
\mu \frac{d}{d\mu} g(\mu;\mu)
\nonumber\\
&& \quad
=
\left.
  {\cal M} \frac{\partial}{\partial {\cal M}} g({\cal M} ;\mu)
\right\vert_{{\cal M} = \mu}
+
\left.
  \mu \frac{\partial}{\partial \mu} g({\cal M} ;\mu)
\right\vert_{{\cal M} = \mu}
\ ,
\label{RGE-g2}
\end{eqnarray}
where the first term in the right-hand-side reproduces
Eq.~(\ref{RGE-mu}) and the second term appears due to the
``intrinsic'' density dependence. Note that $g=0$ is a fixed point
when the second term is neglected (this follows from
(\ref{RGE-mu})), and the presence of the second term makes $g=0$
be no longer the fixed point of Eq.~(\ref{RGE-g2})~\footnote{The
condition $g(\mu_c;\mu_c)=0$ follows from the fixed point of the
RGE in $\M$, but it is not a fixed point of the RGE in $\mu$.}.
The second term can be determined from QCD through the Wilsonian
matching. However, we do not presently have reliable estimate of
the $\mu$ dependence of the QCD correlators. Analyzing the $\mu$
dependence away from the critical density in detail requires a lot
more work, so we relegate this issue to a later publication.

The Wilsonian matching of the correlators at
$\Lambda=\Lambda_\chi$ allows one to see how the $\rho$ mass
scales very near the critical density (or temperature). For this
purpose, it suffices to look at the intrinsic density dependence
of $M_\rho$. We find that close to $\mu_c$
 \ba
M_\rho^2 (\Lambda;\mu)\sim \frac{\la\bar{q}q (\mu)\ra^2}{F_\pi^2
(\Lambda;\mu) \Lambda^2}
 \ea
which implies that
 \ba
\frac{m_\rho^\star}{m_\rho}\sim
\frac{\la\bar{q}{q}\ra^\star}{\la\bar{q}{q}\ra}.\label{scaling}
 \ea
Here the star denotes density dependence.  Note that Equation
(\ref{scaling}) is consistent with the ``Nambu scaling" or more
generally with sigma-model scaling. How this scaling fares with
nature is discussed in \cite{BR:PR2001}.

The following observations can be drawn from this work.
\begin{itemize}
\item
The $parameters$ of BR scaling Lagrangian~\cite{BRscaling} could
be identified with those of HLS Lagrangian that are
Wilsonian-matched at the matching scale and flow to the fixed
point $(g,a,m_q)=(0,1,0)$ with increasing density. An interesting
question here is: How is the BR scaling which is related to Landau
Fermi liquid interaction at normal matter density (reviewed in
\cite{BR:PR2001}) interpreted in terms of the HLS flow?
\item It seems plausible that the density-dependent vector
meson mass that arises via a Higgs mechanism in the color-flavor
locking discussed in \cite{BR:Berkeley,BR:PR2001} refers to
$M_\rho (\mu)$, namely the part that reflects what was interpreted
in \cite{BR:PR2001} as ``sliding vacuum." This is the
Lorentz-invariant piece of the mass. The physical pole mass should
contain also the dense loop corrections that take into account the
velocity $v\neq 1$.
\item
The ``intrinsic" density dependence that is governed by the
Wilsonian matching with QCD and the VM fixed points, as in the
case of ``intrinsic" temperature dependence discussed by Harada
and Sasaki~\cite{HS:T}, is mostly, if not completely, missing in
most of the model descriptions published in the literature. For
instance, the prescription of replacing $m_V$ by $m_V^\star$ near
chiral restoration in the Rapp-Wambach approach as described in
\cite{BLRRW,KRBR} -- which seemed ad hoc at the time those papers
were written -- reflects what is lacking in the Rapp-Wambach
formulation near chiral restoration and may be justified by the
``sliding vacuum'' effect.
\item The notion of density dependence of the cutoff and the
Morley-Kislinger procedure invoked here in the low density region
imply that the cutoff used
in effective field theories should drop as density is increased.
This supports the early suggestion of Adami and Brown~\cite{GEB}
that the cutoff in in-medium NJL model should be density
dependent.
\end{itemize}

To summarize, we have shown that the vector manifestation (VM) is
realized in dense matter at the chiral restoration with the vector
meson mass $m_\rho$ going to zero in the chiral limit. Thus the VM
is $universal$ in the sense that it occurs at $N_f^c$ for
$T=\mu=0$, at $T_c$ for $N_f < N_F^c$ and $\mu=0$ and at $\mu_c$
for $T=0$ and $N_f <N_f^c$. This scenario is characterized by the
common feature that at the chiral transition, the longitudinal
component of the $\rho$ meson joins the pion into a degenerate
multiplet, a scenario which differs from the standard sigma model
scenario. Since the gauge coupling constant $g$ is to go to zero
near the critical point, the dropping-mass vector meson will
become sharper with vanishing width as suggested in \cite{HY:VM},
a phenomenon which cannot be accessed by a strong-coupling theory
valid at low density that involves an expanding width~\cite{rapp}.

\subsection*{Acknowledgments}

We are indebted to Gerry Brown for useful comments. We are also
grateful for discussions with Hyun Kyu Lee and Koichi Yamawaki.
Two of us (MH and MR) acknowledge the hospitality of Korea
Institute for Advanced Study where this work was done. The work of
MH is supported in part by Grant-in-Aid for Scientific Research
(A)\#12740144 and that of YK by the BK21 project of the Ministry
of Education.

\appendix

\section{Effects of Lorentz symmetry breaking}

\setcounter{equation}{0}
\renewcommand{\theequation}{\mbox{A.\arabic{equation}}}

The discussion in the main text was made without explicit
consideration of the effect of Lorentz symmetry breaking inherent
in dense medium. In this Appendix, we examine if and how the
Lorentz symmetry breaking affects the conditions (\ref{g a
z12:VMT}) at $n_c$. Modulo the intrinsic density dependence of the
bare parameters, we obtain the general conditions -- which are an
extension of the conditions in Eq.~(\ref{g a z12:VMT}) to a system
without Lorentz invariance -- by simply requiring that the
axial-vector and vector-current correlators in the HLS agree,
$G_{V(HLS)}^{T,L}=G_{A(HLS)}^{T,L}$, at $n_c$ without considering
the matching to OPE in QCD. A more complete analysis that includes
the matching to OPE in QCD with Lorentz symmetry breaking which is
needed to describe processes away from the critical point will be
reported elsewhere.

\subsection{Polarization tensors}

We start by summarizing the polarization tensors used here and
also in the succeeding Appendices. In hot and/or dense matter, the
polarization tensors are no longer restricted to be Lorentz
covariant, but should be $O(3)$ covariant. Thus we need four
independent symmetric $O(3)$ tensors. Here we adopt the following
form~\cite{ka}:
\begin{eqnarray}
  P_T^{\mu\nu}
&\equiv&
  g^\mu_i
  \left(
    \delta_{ij} - \frac{\vec{p}_i \vec{p}_j}{ \bar{p} }
  \right)
  g_j^\nu
\nonumber\\
&=&
  \left( g^{\mu\alpha} - u^\mu u^\alpha \right)
  \left(
    - g_{\alpha\beta} - \frac{p^\alpha p^\beta}{\bar{p}^2}
  \right)
  \left( g^{\beta\nu} - u^\beta u^\nu \right)
\ ,
\nonumber\\
  P_L^{\mu\nu}
&\equiv&
  - \left( g^{\mu\nu} - \frac{p^\mu p^\nu}{p^2} \right)
  - P_T^{\mu\nu}
\nonumber\\
&=&
  \left( g^{\mu\alpha} - \frac{p^\mu p^\alpha}{p^2} \right)
  u_\alpha
  \frac{p^2}{\bar{p}^2}
  u_\beta
  \left( g^{\beta\nu} - \frac{p^\beta p^\nu}{p^2} \right)
\ ,
\nonumber\\
  P_C^{\mu\nu}
&\equiv&
  \frac{1}{\sqrt{2} \bar{p}}
  \left[
    \left( g^{\mu\alpha} - \frac{p^\mu p^\alpha}{p^2} \right)
    u_\alpha p^\nu\right.\nonumber\\
    && + \left.
    p^\mu u_\beta
    \left( g^{\beta\nu} - \frac{p^\beta p^\nu}{p^2} \right)
  \right]
\ ,
\nonumber\\
  P_D^{\mu\nu}
&\equiv&
  \frac{p^\mu p^\nu}{p^2}
\ ,
\label{pols}
\end{eqnarray}
where $p^\mu =(p_0 , \vec{p})$ is the four-momentum and $\bar{p}
\equiv \vert \vec{p} \vert$. The rest frame of the medium is
indicated by
\begin{equation}
u^\mu = (1,\vec{0})
\ .
\label{def u}
\end{equation}
The polarization tensors satisfy the following
identities~\cite{ka}:
\begin{eqnarray}
&&
  P_T^{\mu\nu}P_{T\mu\nu}=2,~g_{\mu\nu} P_T^{\mu\nu}=-2
\ , \quad
  u_\mu P_T^{\mu\nu}=0
\ ,
\nonumber\\
&&
  P_L^{\mu\nu}P_{L\mu\nu}=1,~g_{\mu\nu} P_L^{\mu\nu}=-1
\ , \quad
  u_\mu u_\nu P_L^{\mu\nu}=\frac{\bar p^2}{p^2}
\ ,
\nonumber\\
&&
  P_C^{\mu\nu}P_{C\mu\nu}=-1
\ ,\quad
  g_{\mu\nu} P_C^{\mu\nu}=0
\ ,
\nonumber\\
&&
  u_\mu u_\nu P_C^{\mu\nu}=-\sqrt{2}\frac{p_0\bar p}{p^2}
\ .
\label{relation:pol}
\end{eqnarray}

\subsection{Axial-vector correlator}

As argued in Section 5 of Ref.~\cite{BR:PRep}, the vector
correlator receives at the chiral restoration point an important
contribution from quasiquark loop diagrams. Such a contribution
cannot in general be expressed by a local effective Lagrangian in
which quasiquarks are absent. In the present work, we are
considering the HLS model that includes the quasiquarks as
explicit degrees of freedom near the critical point. Therefore we
consider it reasonable to assume that the {\it bare} HLS theory we
are concerned with can be expressed by a {\it local} Lagrangian
with the nonlocal quasiquark contribution appearing at the later
stage of Wilsonian decimation.

The bare HLS Lagrangian in hot and/or dense matter is generally
expected to include the effect of Lorentz non-invariance. The
Lagrangian density valid to ${\cal O}(p^4)$ relevant to the
axial-vector current correlator can be written as
\begin{eqnarray}
&&
{\cal L}_{(A)}
=
\biggl[
  (F_{\pi,{\rm bare}}^t)^2 u_\mu u_\nu
\nonumber\\
&& \qquad
  +
  F_{\pi,{\rm bare}}^t F_{\pi,{\rm bare}}^s
    \left( g_{\mu\nu} - u_\mu u_\nu \right)
\biggr]
\mbox{tr}
\left[
  \hat{\alpha}_\perp^\mu \hat{\alpha}_\perp^\nu
\right]
\nonumber\\
&& \quad
+
\biggl[
  2 z^L_{2,{\rm bare}} \, u_\mu u_\alpha g_{\nu\beta}
\nonumber\\
&& \qquad
  + z^T_{2,{\rm bare}}
  \left(
    g_{\mu\alpha} g_{\nu\beta}
   - 2 u_\mu u_\alpha g_{\nu\beta}
  \right)
\biggr]
\,
\mbox{tr}
\left[ \widehat{\cal A}^{\mu\nu} \widehat{\cal A}^{\alpha\beta} \right]
\ ,
\label{Lag:1}
\end{eqnarray}
where $F_{\pi,{\rm bare}}^t$ and $F_{\pi,{\rm bare}}^s$ denote the
{\it bare} parameters associated with the temporal and spatial
pion decay constants. The rest frame of the medium is specified by
$u^\mu$ as in Eq.~(\ref{def u}). The parameters $z^L_{2,{\rm
bare}}$ and $z^T_{2,{\rm bare}}$ correspond in medium to the
vacuum parameter $z_{2,{\rm bare}}$~\cite{Tanabashi,HY:matching}
at $T = \mu =0$, and $\widehat{\cal A}^{\mu\nu}$ is defined by
\begin{equation}
\widehat{\cal A}_{\mu\nu} \equiv \frac{1}{2}
\left[
  \xi_{\rm R} {\cal R}_{\mu\nu} \xi_{\rm R}^\dag
  -
  \xi_{\rm L} {\cal L}_{\mu\nu} \xi_{\rm L}^\dag
\right]
\ ,
\end{equation}
where ${\cal R}_{\mu\nu}$ and ${\cal L}_{\mu\nu}$ are the
field-strength tensors of the external gauge fields ${\cal R}_\mu$
and ${\cal L}_\mu$:
\begin{eqnarray}
&& {\cal L}_{\mu\nu} = \partial_\mu {\cal L}_\nu -
\partial_\nu {\cal L}_\mu -
i \left[ {\cal L}_\mu \,,\, {\cal L}_\nu \right] \ ,
\nonumber\\
&&
{\cal R}_{\mu\nu} = \partial_\mu {\cal R}_\nu -
\partial_\nu {\cal R}_\mu -
i \left[ {\cal R}_\mu \,,\, {\cal R}_\nu \right] \ .
\end{eqnarray}
Note
that in the bare theory $\hat{\alpha}_\perp^\mu$ is expanded as
\begin{eqnarray}
\hat{\alpha}_\perp^\mu = {\cal A}^\mu +
\frac{\partial_\mu \pi}{F_{\pi,{\rm bare}}^t} + \cdots \
\ ,
\end{eqnarray}
where
${\cal A}_\mu = ({\cal R}_\mu - {\cal L}_\mu)/2$.

We define the axial-vector current correlator $G_A^{\mu\nu}(p)$ by
\begin{eqnarray}
&&
i \int d^4 x e^{i p x}
\left\langle 0 \left\vert T\, J_{5\mu}^a (x) J_{5\nu}^b (0)
\right\vert 0 \right\rangle
= \delta^{ab} G_A^{\mu\nu}(p)
\ ,
\label{def GA}
\end{eqnarray}
and decompose it into
\begin{eqnarray}
G_A^{\mu\nu}(p) = P_L^{\mu\nu} G_A^L(p) + P_T^{\mu\nu} G_A^T(p) \
.\label{decompose}
\end{eqnarray}
It follows from the bare HLS Lagrangian, Eq.~(\ref{Lag:1}), and
Fig.~\ref{fpt-1} that
\begin{eqnarray}
&&
G_{A{\rm(HLS)}}^{\mu\nu}(p)
=
\frac{ \widetilde{\Gamma}_{A,{\rm bare}}^\mu
       \widetilde{\Gamma}_{A,{\rm bare}}^\nu}%
{ - [ p_0^2 - v_{{\rm bare}}^2 \bar{p}^2 ] }
\nonumber\\
&& \quad
{}+
\left[
  (F_{\pi,{\rm bare}}^t)^2 u^\mu u^\nu
  +
  F_{\pi,{\rm bare}}^t F_{\pi,{\rm bare}}^s
  \left( g^{\mu\nu} - u^\mu u^\nu \right)
\right]
\nonumber\\
&& \quad
{}
- 2 z^L_{2,{\rm bare}}\, p^2 P_L^{\mu\nu}
\nonumber\\
&& \quad
{}- 2 \left(
  z^L_{2,{\rm bare}}\, p_0^2 - z^T_{2,{\rm bare}}\, \bar{p}^2
\right)
P_T^{\mu\nu}
\ ,
\label{GA HLS}
\end{eqnarray}
where $v_{{\rm bare}}$ is the bare pion velocity related to
$F_{\pi,{\rm bare}}^t$ and $F_{\pi,{\rm bare}}^s$ by
\begin{equation}
v_{{\rm bare}}^2
= \frac{F_{\pi, {\rm bare}}^s}{F_{\pi, {\rm bare}}^t}
\ ,
\end{equation}
and
\begin{eqnarray}
\widetilde{\Gamma}_{A,{\rm bare}}^\mu
&\equiv&
\biggl[
  F_{\pi,{\rm bare}}^t u_\mu u_\alpha
\nonumber\\
&&
  {} +
  F_{\pi,{\rm bare}}^s \left( g_{\mu\alpha} - u_\mu u_\alpha \right)
\biggr]
p^\alpha
\ .
\end{eqnarray}
To obtain the $P_L^{\mu\nu} $ and $P_T^{\mu\nu} $ terms in
Eq.~(\ref{GA HLS}), we have used the following identities:
\begin{eqnarray}
&&
  (u\cdot p)^2 g^{\mu\nu} + p^2 u^\mu u^\nu -
  (u\cdot p) \left( u^\mu p^\nu + p^\mu u^\nu \right)
\nonumber\\
&& \quad
=
  - p^2 P_L^{\mu\nu} - p_0^2 P_T^{\mu\nu}
\nonumber\\
&& \quad
=
  \left( p^2 g^{\mu\nu} - p^\mu p^\nu \right)
  - \bar{p}^2 P_T^{\mu\nu}
\ .
\end{eqnarray}
Now, by using the identities Eq.~(\ref{relation:pol}), we obtain
from Eq.~(\ref{decompose}) and Eq.~(\ref{GA HLS}),
\begin{eqnarray}
&&
G_{A{\rm(HLS)}}^L(p)
=
\frac{ p^2 F_{\pi,{\rm bare}}^t F_{\pi,_{\rm bare}}^s }{
   - [ p_0^2 - v_{\rm bare}^2 \bar{p}^2 ] }
-2p^2z^L_{2,\rm bare}
\label{gal} \\
&&
G_{A{\rm(HLS)}}^T(p)
=
-F_{\pi,\rm bare}^tF_{\pi,\rm bare}^s
\nonumber\\
&& \qquad{}
  - 2 \left(
    p_0^2 z_{2,{\rm bare}}^L  - \bar{p}^2 z_{2,{\rm bare}}^T
  \right)
\ .
\label{gat}
\ea

\begin{figure}[htb]
\centerline{\epsfig{file=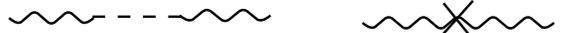,width=7.3cm}} \vskip 0.3cm
\caption{\small Tree-level contributions to the
 axial-vector (vector) current correlator.
The dashed line denotes pion for axial-vector correlator or $\rho$ meson
for vector correlator.}\label{fpt-1}
\end{figure}

\subsection{Vector correlator}

The {\it bare} Lagrangian density valid to ${\cal O}(p^4)$
relevant to the vector correlator  is
\begin{eqnarray}
&&
{\cal L}_{(V)}
=
\biggl[
  (F_{\sigma,{\rm bare}}^t)^2 u_\mu u_\nu
\nonumber\\
&& \qquad\quad
  {}+
  F_{\sigma,{\rm bare}}^t F_{\sigma,{\rm bare}}^s
    \left( g_{\mu\nu} - u_\mu u_\nu \right)
\biggr]
\mbox{tr}
\left[
  \hat{\alpha}_\parallel^\mu \hat{\alpha}_\parallel^\nu
\right]
\nonumber\\
&& \quad
{} +
\biggl[
  2 z^L_{1,{\rm bare}} \, u_\mu u_\alpha g_{\nu\beta}
\nonumber\\
&& \qquad\quad
  {}  + z^T_{1,{\rm bare}}
  \left(
    g_{\mu\alpha} g_{\nu\beta}
   - 2 u_\mu u_\alpha g_{\nu\beta}
  \right)
\biggr]
\,
\mbox{tr}
\left[ \widehat{\cal V}^{\mu\nu} \widehat{\cal V}^{\alpha\beta} \right]
\nonumber\\
&& \quad
{} +
\biggl[
  2 z^L_{3,{\rm bare}} \, u_\mu u_\alpha g_{\nu\beta}
\nonumber\\
&& \qquad\quad
  {} + z^T_{3,{\rm bare}}
  \left(
    g_{\mu\alpha} g_{\nu\beta}
   - 2 u_\mu u_\alpha g_{\nu\beta}
  \right)
\biggr]
\,
\mbox{tr}
\left[ V^{\mu\nu} \widehat{\cal V}^{\alpha\beta} \right]
\nonumber\\
&& \quad
{} +
\Biggl[
  - \frac{1}{ g_{L,{\rm bare}}^2 } \, u_\mu u_\alpha g_{\nu\beta}
\nonumber\\
&& \qquad\quad
  - \frac{1}{ 2 g_{T,{\rm bare}}^2 }
  \left(
    g_{\mu\alpha} g_{\nu\beta}
   - 2 u_\mu u_\alpha g_{\nu\beta}
  \right)
\Biggr]
\,
\mbox{tr}
\left[ V^{\mu\nu} V^{\alpha\beta} \right]
\ ,
\nonumber\\
\label{Lag:V}
\end{eqnarray}
where
$\widehat{\cal V}^{\mu\nu}$ is defined
by
\begin{equation}
\widehat{\cal V}_{\mu\nu} \equiv \frac{1}{2}
\left[
  \xi_{\rm R} {\cal R}_{\mu\nu} \xi_{\rm R}^\dag
  +
  \xi_{\rm L} {\cal L}_{\mu\nu} \xi_{\rm L}^\dag
\right]
\ .
\end{equation}
$F_{\sigma,{\rm bare}}^t$ and $F_{\sigma,{\rm bare}}^s$ denote the
bare parameters associated with the temporal and spatial
components of the decay constants of the $\sigma$ (i.e., the
longitudinal $\rho$). We define the matter-extension of the
parameter $a$ as
\begin{equation}
a^t \equiv
\left( \frac{F_{\sigma,{\rm bare}}^t}{F_{\pi,{\rm bare}}^t} \right)^2
\ , \quad
a^s \equiv
\left( \frac{F_{\sigma,{\rm bare}}^s}{F_{\pi,{\rm bare}}^s} \right)^2
\ .
\end{equation}
$z_{1,{\rm bare}}^L$ and $z_{1,{\rm bare}}^T$ in Eq.~(\ref{Lag:V})
correspond in medium to $z_{1,{\rm bare}}$, $z_{3,{\rm bare}}^L$
and $z_{3,{\rm bare}}^T$ to $z_{3,{\rm bare}}$ and $g_{L,{\rm
bare}}$ and $g_{T,{\rm bare}}$ to $g_{{\rm bare}}$.

Let us first obtain a general form of the vector current
correlator $G_V^{\mu\nu}$, defined in an analogous way to
$G_A^{\mu\nu}$ in Eq.~(\ref{def GA}). By using the vector meson
propagator $i D_{\mu\nu}$, the $V$-${\cal V}$ two-point function
$\Pi_{V\parallel}$ and the ${\cal V}$-${\cal V}$ two-point
function $\Pi_{\parallel}$, we can express $G_V^{\mu\nu}$ in HLS
theory as
\begin{eqnarray}
G_{V{\rm(HLS)}}^{\mu\nu} = \Pi_{V\parallel}^{\mu\alpha} \,
i D_{\alpha\beta} \, \Pi_{V\parallel}^{\beta\nu}
+
\Pi_{\parallel}^{\mu\nu}
\ .
\label{GV form}
\end{eqnarray}
The vector meson propagator $i D_{\mu\nu}$ is related to the
$V$-$V$ two-point function $\Pi_V^{\mu\nu}$ by
\begin{equation}
i \left( D^{-1} \right)^{\mu\nu} = \Pi_V^{\mu\nu} \ .
\label{Dinv PiV}
\end{equation}
It is convenient to decompose
$\Pi_V^{\mu\nu}$ into the following
four independent pieces:
\begin{eqnarray}
\Pi_V^{\mu\nu} = g^{\mu\nu} \Pi_V^S + P_L^{\mu\nu} \Pi_V^L +
P_T^{\mu\nu} \Pi_V^T + P_C^{\mu\nu} \Pi_V^C \ , \label{decomp:PiV}
\end{eqnarray}
in terms of which the vector meson propagator is given by
\begin{eqnarray}
  - i D^{\mu\nu}
&=&
  P_L^{\mu\nu}
  \frac{\Pi_V^S}{
    \Pi_V^S ( \Pi_V^L - \Pi_V^S ) - (\Pi_V^C)^2/2
  }
\nonumber\\
&&
  {}+
  P_T^{\mu\nu}
  \frac{1}{\Pi_V^T - \Pi_V^S}
\nonumber\\
&&
  {}+
  P_C^{\mu\nu}
  \frac{\Pi_V^C}{
    \Pi_V^S ( \Pi_V^L - \Pi_V^S ) - (\Pi_V^C)^2/2
  }\nonumber\\
&&
  {}+
  P_D^{\mu\nu}
  \frac{\Pi_V^L - \Pi_V^S}{
    \Pi_V^S ( \Pi_V^L - \Pi_V^S ) - (\Pi_V^C)^2/2
  }
\ . \label{prop:form1}
\end{eqnarray}
Similarly to $\Pi_V^{\mu\nu}$,
the two-point functions
$\Pi_{V\parallel}$ and $\Pi_{\parallel}$
can be decomposed as
\begin{eqnarray}
&&
\Pi_{V\parallel}^{\mu\nu}
= g^{\mu\nu} \Pi_{V\parallel}^S + P_L^{\mu\nu} \Pi_{V\parallel}^L
+ P_T^{\mu\nu} \Pi_{V\parallel}^T + P_C^{\mu\nu} \Pi_{V\parallel}^C
\ ,
\nonumber
\\
&&
\Pi_{\parallel}^{\mu\nu}
= g^{\mu\nu} \Pi_{\parallel}^S + P_L^{\mu\nu} \Pi_{\parallel}^L
+ P_T^{\mu\nu} \Pi_{\parallel}^T + P_C^{\mu\nu} \Pi_{\parallel}^C
\ .
\label{decomp:Pip}
\end{eqnarray}
With Eqs.~(\ref{prop:form1}) and (\ref{decomp:Pip}),
$G_{V{\rm(HLS)}}^{\mu\nu}$ reads
\begin{eqnarray}
G_{V{\rm(HLS)}}^{\mu\nu}
&=&
P_T^{\mu\nu} G_{V{\rm(HLS)}}^T
+ P_L^{\mu\nu} G_{V{\rm(HLS)}}^L
\nonumber\\
&&
{}+ P_C^{\mu\nu} G_{V{\rm(HLS)}}^C
+ P_D^{\mu\nu} G_{V{\rm(HLS)}}^D
\ ,
\end{eqnarray}
where
\begin{eqnarray}
&&
G_{V{\rm(HLS)}}^T
=
\frac{
  - \left( \Pi_{V\parallel}^T - \Pi_{V\parallel}^S \right)^2
}{
  \Pi_V^T - \Pi_V^S
}
+ \Pi_{\parallel}^T - \Pi_{\parallel}^S
\ ,
\nonumber\\
&&
G_{V{\rm(HLS)}}^L
=
\frac{1}{
  \Pi_V^S ( \Pi_V^L - \Pi_V^S ) - (\Pi_V^C)^2/2
}
\nonumber\\
&& \quad
\times
\Biggl[
  - \Pi_V^S \left( \Pi_{V\parallel}^L - \Pi_{V\parallel}^S \right)^2
  {}+ \Pi_V^C \Pi_{V\parallel}^C
    \left( \Pi_{V\parallel}^L - \Pi_{V\parallel}^S \right)
\nonumber\\
&& \qquad
  {} - \frac{1}{2} \left( \Pi_V^L - \Pi_V^S \right)
  \left( \Pi_{V\parallel}^C \right)^2
\Biggr]
+
\left( \Pi_{\parallel}^L - \Pi_{\parallel}^S \right)
\ ,
\nonumber\\
&&
G_{V{\rm(HLS)}}^C
=
\frac{1}{
  \Pi_V^S ( \Pi_V^L - \Pi_V^S ) - (\Pi_V^C)^2/2
}
\nonumber\\
&& \quad
\times
\Biggl[
  - \Pi_V^C
  \left\{
    \left( \Pi_{V\parallel}^S \right)^2
    - \Pi_{V\parallel}^S \Pi_{V\parallel}^L
    - \left( \Pi_{V\parallel}^C \right)^2 / 2
  \right\}
\nonumber\\
&& \qquad
  {}- \Pi_V^S \Pi_{V\parallel}^C
    \left( \Pi_{V\parallel}^L - \Pi_{V\parallel}^S \right)
\nonumber\\
&& \qquad
  {}- \left( \Pi_V^L - \Pi_V^S \right)
  \Pi_{V\parallel}^S \Pi_{V\parallel}^C
\Biggr]
+
\Pi_{\parallel}^C
\ ,
\nonumber\\
&&
G_{V{\rm(HLS)}}^D
=
\frac{1}{
  \Pi_V^S ( \Pi_V^L - \Pi_V^S ) - (\Pi_V^C)^2/2
}
\nonumber\\
&& \quad
\times
\Biggl[
  \Pi_V^C \Pi_{V\parallel}^S \Pi_{V\parallel}^C
  - \Pi_V^S \left( \Pi_{V\parallel}^C \right)^2/2
\nonumber\\
&& \qquad
  {}- \left( \Pi_V^L - \Pi_V^S \right)
  \left( \Pi_{V\parallel}^S \right)^2
\Biggr]
+
\Pi_{\parallel}^S
\ .
\end{eqnarray}
The requirement for the current conservation is that
$G_{V{\rm(HLS)}}^C$ and $G_{V{\rm(HLS)}}^D$ vanish. We can easily
see that
\begin{equation}
G_{V{\rm(HLS)}}^C = C_{V{\rm(HLS)}}^D = 0 \ ,
\end{equation}
when the following conditions are satisfied~\footnote{%
  Two conditions in Eq.~(\ref{cond:SC})
  are actually satisfied by the contributions obtained from
  the bare Lagrangian in Eq.~(\ref{Lag:V}).
}
\begin{eqnarray}
&&
  \Pi_V^S = \Pi_\parallel^S = - \Pi_{V\parallel}^S
\ ,
\nonumber
\\
&&
  \Pi_V^C = \Pi_\parallel^C = - \Pi_{V\parallel}^C
\ .
\label{cond:SC}
\end{eqnarray}
Then, $G_{V{\rm(HLS)}}^T$ and $G_{V{\rm(HLS)}}^L$ can be rewritten
as
\begin{eqnarray}
&&
G_{V{\rm(HLS)}}^T
=
\frac{
  - \left( \Pi_{V\parallel}^T + \Pi_V^S \right)^2
}{
  \Pi_V^T - \Pi_V^S
}
+ \Pi_{\parallel}^T - \Pi_V^S
\ ,
\nonumber\\
&&
G_{V{\rm(HLS)}}^L
=
\frac{1}{
  \Pi_V^S ( \Pi_V^L - \Pi_V^S ) - (\Pi_V^C)^2/2
}
\nonumber\\
&& \quad
\times
\Biggl[
  - \Pi_V^S \left( \Pi_{V\parallel}^L + \Pi_V^S \right)^2
\nonumber\\
&& \qquad
  {}- \frac{1}{2} \left( \Pi_V^C \right)^2
  \left( \Pi_V^L + \Pi_V^S + 2 \Pi_{V\parallel}^L \right)
\Biggr]
\nonumber\\
&& \quad
{}+
\left( \Pi_{\parallel}^L - \Pi_V^S \right)
\ .
\end{eqnarray}

Now, using the bare Lagrangian (\ref{Lag:V}), we find (here and
below, the subscript ``bare" is omitted to simplify writing)
\begin{eqnarray}
&&
\Pi_V^S
= \Pi_\parallel^S = - \Pi_{V\parallel}^S
=
\frac{p_0^2}{p^2} \left( F_\sigma^t \right)^2
-
\frac{\bar{p}^2}{p^2} \left( F_\sigma^t F_\sigma^s \right)
\ ,
\nonumber\\
&& \Pi_V^C = \Pi_\parallel^C = - \Pi_{V\parallel}^C = \sqrt{2}
\frac{p_0 \bar{p}}{p^2} F_\sigma^t \left( F_\sigma^t - F_\sigma^s
\right) \ ,\label{S C expressions}
\end{eqnarray}
and
\begin{eqnarray}
\Pi_V^L
&=&
\frac{p_0^2 + \bar{p}^2 }{p^2} F_\sigma^t
\left( F_\sigma^t - F_\sigma^s \right)
+ \frac{p^2}{g_L^2}
\ ,
\nonumber\\
\Pi_V^T
&=&
\frac{p_0^2 }{p^2} F_\sigma^t
\left( F_\sigma^t - F_\sigma^s \right)
+ \frac{p_0^2}{g_L^2}
  - \frac{\bar{p}^2}{g_T^2}
\ ,
\nonumber\\
\Pi_{V\parallel}^L
&=&
-
\frac{p_0^2 + \bar{p}^2 }{p^2} F_\sigma^t
\left( F_\sigma^t - F_\sigma^s \right)
- p^2 z_{3}^L
\ ,
\nonumber\\
\Pi_{V\parallel}^T
&=&
-
\frac{p_0^2 }{p^2} F_\sigma^t
\left( F_\sigma^t - F_\sigma^s \right)
 - \left( p_0^2 z_{3}^L - \bar{p}^2 z_{3}^T \right)
\ ,
\nonumber\\
\Pi_{\parallel}^L
&=&
\frac{p_0^2 + \bar{p}^2 }{p^2} F_\sigma^t
\left( F_\sigma^t - F_\sigma^s \right)
- 2 p^2 z_{1}^L
\ ,
\nonumber\\
\Pi_{\parallel}^T
&=&
\frac{p_0^2 }{p^2} F_\sigma^t
\left( F_\sigma^t - F_\sigma^s \right)
- 2 \left( p_0^2 z_{1}^L - \bar{p}^2 z_{1}^T \right)
\ .
\nonumber\\
\label{L T expressions}
\end{eqnarray}
Finally the results are
\begin{eqnarray}
&&
G_{V{\rm(HLS)}}^L
=
\frac{
  p^2 \, F_\sigma^t F_\sigma^s
  \left( 1 - 2 g_L^2 z_3^L \right)
}{
  -
  \left[
    p_0^2 - \left( F_\sigma^s / F_\sigma^t \right) \bar{p}^2
    - M_v^2
  \right]
}
\nonumber\\
&& \quad
{}- 2 p^2 z_{1}^L
+ {\cal O}(p^4)
\ ,
\nonumber\\
&&
G_{V{\rm(HLS)}}^T
=
\frac{
  F_\sigma^t F_\sigma^s
}{
  - \left[
    p_0^2 - \left(g_L^2/g_T^2\right) \bar{p}^2 - M_v^2
  \right]
}
\nonumber\\
&& \quad
\times
  \left[
    p_0^2 - \left(g_L^2/g_T^2\right) \bar{p}^2
    - 2 g_L^2 \left( p_0^2 z_{3}^L - \bar{p}^2 z_{3}^T \right)
  \right]
\nonumber\\
&& \quad
{}- 2 \left( p_0^2 z_{1}^L - \bar{p}^2 z_{1}^T \right)
+ {\cal O}(p^4)
\ ,
\end{eqnarray}
in which we have dropped terms proportional to $\left( p^2 z_{3}^L
\right)^2$ and $\left( p_0^2 z_{3}^L - \bar{p}^2 z_{3}^T
\right)^2$ which are of higher order in the present counting
scheme together with other ${\cal O}(p^4)$ terms that require
${\cal O}(p^6)$ Lagrangian density. In the above expressions $M_v$
is the bare mass at rest frame:
\begin{equation}
M_v^2 \equiv g_L^2 F_\sigma^t F_\sigma^s \ .
\end{equation}
It should be noticed that, at rest frame, $G_{V{\rm(HLS)}}^L$ is
equal to $G_{V{\rm(HLS)}}^T$ and that both the longitudinal and
transverse modes of the vector meson have the same bare mass.

\subsection{The equality
  $G_{V{\rm(HLS)}}^{T,L}=G_{A{\rm(HLS)}}^{T,L}$ at $n_c$}

At the chiral phase transition point, the axial-vector and vector
current correlators must agree with each other: $G_{A{\rm(HLS)}}^L
= G_{V{\rm(HLS)}}^L$ and $G_{A{\rm(HLS)}}^T = G_{V{\rm(HLS)}}^T$.
Imposing this condition, we obtain
\begin{eqnarray}
&&
\frac{ p^2 F_{\pi}^t F_{\pi}^s }{
   - [ p_0^2 - \left( F_\pi^s/F_\pi^t \right) \bar{p}^2 ] }
- 2 p^2 z_{2}^L
\nonumber\\
&& \qquad
=
\frac{
  p^2 \, F_\sigma^t F_\sigma^s
  \left( 1 - 2 g_L^2 z_3^L \right)
}{
  -
  \left[
    p_0^2 - \left( F_\sigma^s / F_\sigma^t \right) \bar{p}^2
    - M_L^2
  \right]
}
- 2 p^2 z_{1}^L
\ ,
\\
&&
- F_{\pi}^t F_{\pi}^s
- 2 \left(
  p_0^2 z_{2}^L  - \bar{p}^2 z_{2}^T
\right)
\nonumber\\
&& \qquad
=
\frac{
  F_\sigma^t F_\sigma^s
  \left[
    p_0^2 - \left(g_L^2/g_T^2\right) \bar{p}^2
    - 2 g_L^2 \left( p_0^2 z_{3}^L - \bar{p}^2 z_{3}^T \right)
  \right]
}{
  - \left[
    p_0^2 - \left(g_L^2/g_T^2\right) \bar{p}^2 - M_L^2
  \right]
}
\nonumber\\
&& \qquad\quad
{}- 2 \left( p_0^2 z_{1}^L - \bar{p}^2 z_{1}^T \right)
\ .
\end{eqnarray}
Now, we see that the above equalities are satisfied for any values
of $p_0$ and $\bar{p}$ around the matching scale only if the
following conditions are met:
\begin{eqnarray}
&&
  a^t = \left( \frac{F_\sigma^t}{F_\pi^t} \right)^2 = 1
\ ,
\quad
  a^s = \left( \frac{F_\sigma^s}{F_\pi^s} \right)^2 = 1
\ ,
\nonumber\\
&&
  g_L = 0
\ ,
  \quad
  g_T = 0
\ ,
\nonumber\\
&&
  z_2^L = z_1^L \ ,
  \quad
  z_2^T = z_1^T
\ .
\end{eqnarray}
These conditions are the Lorentz non-invariant version of the
vector-manifestation conditions, Eq.~(\ref{g a z12:VMT}),
discussed in the main text.

\section{Pion Decay Constant}
\label{app:fpi}
\setcounter{equation}{0}
\renewcommand{\theequation}{\mbox{B.\arabic{equation}}}

In this Appendix, we show how the dense hadronic corrections to
the pion decay constant vanish at $\mu=\mu_c$.
In this and next Appendices we neglect possible
effects of Lorentz symmetry
breaking in the bare theory of the HLS:
We assume that
the bare HLS Lagrangian possesses the Lorentz invariance
and that the dominant effects of Lorentz symmetry breaking
come from the quasiquark dense loop correction.
The relevant Lagrangian involving quasiquark
fields at ${\cal O} (p)$ is given in Eq.~(\ref{lagbaryon}).
Here we also include the following ${\cal O} (p^2)$ Lagrangian
 \ba
\delta{\cal L}_{Q(2A)} &=& \bar{\psi}\left\{
  c_{A1} \mbox{tr}
    \left[ \hat{\alpha}_{\perp\mu} \hat{\alpha}_{\perp}^\mu \right]
  + c_{A2}
    \hat{\alpha}_{\perp\mu}
    \hat{\alpha}_{\perp}^\mu\right\}\psi\nonumber\\
  &&+\bar{\psi} c_{A3}
    \left[
      \hat{\alpha}_{\perp\mu} \,,\, \hat{\alpha}_{\perp\nu}
    \right]
    \left[ \gamma^\mu \,,\, \gamma^\nu \right]
\psi \ , \label{Lag:Q2A}
 \ea
where the $c$'s are constants of mass dimension $-1$~\footnote{For
$N_f=2$, the $c_{A1}$-term and $c_{A2}$-term are not independent.
We can set, e.g., $c_{A2}=0$ without loss of generality.}. The
${\cal O}(p^2)$ Lagrangian in Eq.~(\ref{Lag:Q2A}) generates the
${\cal O}(p^5)$ corrections to the first term of the leading order
HLS Lagrangian in Eq.~(\ref{Lagrangian}) at one loop. Since the
relevant diagrams are tadpoles, all the divergent corrections are
proportional to $m_q$. Then the RGE for $F_\pi$ is not changed at
the critical density where $m_q$ vanishes.

Our convention and notations in this and succeeding appendices
are: $p^\mu = ( p_0, \vec{p})$, $\bar{p} \equiv \vert \vec{p}
\vert$, $\omega_M( \bar{p}) \equiv \sqrt{ M^2 + \bar{p}^2 }$ in
free space, $\omega(\bar{p}) \equiv \omega_0(\bar{p}) = \bar{p}$
and for the pion $\tilde{\omega}(\bar{p}) \equiv
\tilde{\omega}_0(\bar{p}) = v(\bar{p}) \bar{p}$ where $v(\bar{p})$
is the pion velocity. The rest frame of the medium will be
indicated by $u^\mu = ( 1, \vec{0} )$ as in Eq.~(\ref{def u}).

In terms of the axial-vector-axial-vector two-point functions
$\Pi^{\mu\nu}_\perp$, the temporal and spatial components of the
pion decay constant are given by
 \ba
 f_\pi^t
&=&
  \frac{1}{\widetilde{F}}
  \left.
    \frac{ u_\mu \Pi^{\mu\nu}_\perp (p_0,\vec{p}) p_\nu }{ p_0 }
  \right\vert_{p_0 = \tilde{\omega}}
\ ,
\nonumber\\
  f_\pi^s
&=&
  \frac{1}{\widetilde{F}}
  \left.
    \frac{
      - p^\alpha ( g_{\alpha\mu} - u_\alpha u_\mu )
      \Pi^{\mu\nu}_\perp (p_0,\vec{p}) p_\nu
    }{ \bar{p}^2 }
  \right\vert_{p_0 = \tilde{\omega}}
\ ,
\label{fpit fpis defs}
 \ea
where $\widetilde{F}$ is the $\pi$ wave function renormalization
constant. According to the analysis of Ref.~\cite{MOR:01} in dense
matter, this $\widetilde{F}$ is nothing but $f_\pi^t$:
\begin{equation}
\widetilde{F} = f_\pi^t \ .
\end{equation}
We wish to compute the $f_\pi^{t,s}$ in HLS including the
quasiquark terms. In HLS, the correlator tensors are
\begin{equation}
  \Pi^{\mu\nu}_\perp (p_0,\vec{p}) =
  g^{\mu\nu} F_\pi^2
  + 2 z_2 \left( g^{\mu\nu} p^2 - p^\mu p^\nu \right)
  + \overline{\Pi}^{\mu\nu}_\perp (p_0,\vec{p})
\ ,
\end{equation}
where $\overline{\Pi}^{\mu\nu}_\perp (p_0,\vec{p})$ denotes the
hadronic dense/thermal corrections we are interested in. On-shell
for the pion, we have
\begin{eqnarray}
  \Pi^{tt}_\perp (\tilde{\omega},\bar{p})
&=&
  F_\pi^2 - 2 z_2 \bar{p}^2
  + \overline{\Pi}^{tt}_\perp (\tilde{\omega},\bar{p})
\ ,
\nonumber\\
  \Pi^{ts}_\perp (\tilde{\omega},\bar{p})
&=&
  2 z_2 v \bar{p}^2
  + \overline{\Pi}^{ts}_\perp (\tilde{\omega},\bar{p})
\ ,
\nonumber\\
  \Pi^{ss}_\perp (\tilde{\omega},\bar{p})
&=&
  - F_\pi^2 - 2 z_2 v^2 \bar{p}^2
  + \overline{\Pi}^{ss}_\perp (\tilde{\omega},\bar{p})
\  \label{Pi perps:onshell}
\end{eqnarray}
where
 \begin{eqnarray} &&
  \Pi^{tt}_\perp (p_0,\bar{p})
  \equiv
  u_\mu \Pi^{\mu\nu}_\perp (p_0,\vec{p}) u_\nu
\ ,
\nonumber\\
&&
  \Pi^{ts}_\perp (p_0,\bar{p})
  \equiv
  \frac{1}{\bar{p}} \,
  u_\mu \Pi^{\mu\nu}_\perp (p_0,\vec{p})
  (g_{\nu\alpha} - u_\nu u_\alpha ) p^\alpha
\nonumber\\
&& \qquad
  =
  \frac{1}{\bar{p}} \,
  p^\alpha (g_{\alpha\mu} - u_\alpha u_\mu )
  \Pi^{\mu\nu}_\perp (p_0,\vec{p})
  u_\nu
  =
  \Pi^{st}_\perp (p_0,\bar{p})
\ ,
\nonumber\\
&&
  \Pi^{ss}_\perp (p_0,\bar{p})
\nonumber\\
&& \quad
  \equiv
  \frac{1}{\bar{p}^2} \,
  p^\alpha (g_{\alpha\mu} - u_\alpha u_\mu )\,
  \Pi^{\mu\nu}_\perp (p_0,\vec{p})\,
  (g_{\nu\beta} - u_\nu u_\beta ) p^\beta \ ,
\nonumber\\
\end{eqnarray}
and $v$ denotes the pion velocity in dense medium
\begin{equation}
v^2 = 1 - \frac{1}{F_\pi^2} \left[
  \overline{\Pi}^{tt}_\perp (\omega,\bar{p})
  + 2 \overline{\Pi}^{ts}_\perp (\omega,\bar{p})
  + \overline{\Pi}^{ss}_\perp (\omega,\bar{p})
\right] \ . \label{v2: form}
\end{equation}
In this expression, we have replaced $\tilde{\omega}$ by $\omega$,
since the difference is of higher order.

Substituting Eq.~(\ref{Pi perps:onshell}) into Eq.~(\ref{fpit fpis
defs}), we obtain
\begin{eqnarray}
&&
  \left[ f_\pi^t \right]^2
  =
  \left[
    \Pi^{tt}_\perp (\tilde{\omega},\bar{p})
    + \frac{1}{v} \Pi^{ts}_\perp (\tilde{\omega},\bar{p})
  \right]_{p_0=\tilde{\omega}}
\nonumber\\
&& \quad
  =
  F_\pi^2 + \overline{\Pi}^{tt}_\perp (\omega,\bar{p})
  + \overline{\Pi}^{ts}_\perp (\omega,\bar{p})
  + {\cal O}(p^4)
\ , \label{fpit2: form}
\\
&&
  f_\pi^t f_\pi^s
  =
  \left[
    - v \Pi^{ts}_\perp (\tilde{\omega},\bar{p})
    - \Pi^{ss}_\perp (\tilde{\omega},\bar{p})
  \right]_{p_0=\tilde{\omega}}
\nonumber\\
&& \quad
  =
  F_\pi^2 - \overline{\Pi}^{ts}_\perp (\omega,\bar{p})
  - \overline{\Pi}^{ss}_\perp (\omega,\bar{p})
  + {\cal O}(p^4)
\ , \label{fpits: form}
\end{eqnarray}
This expression is consistent with the relation $v^2 =
\frac{f_\pi^s}{f_\pi^t}$.

\begin{figure}[htb]
\centerline{\epsfig{file=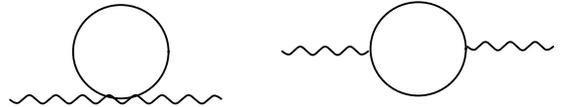,width=7.3cm}} \vskip 0.3cm
\caption{\small The hadronic dense (loop) corrections to the
vector-vector or axial-vector-axial-vector two-point functions.
 The solid lines denote quasiquarks.}\label{fav}
\end{figure}

To compute (\ref{fpit2: form}) and (\ref{fpits: form}), we need to
compute hadronic dense fluctuation (loop) terms given in Fig. 1
with (\ref{lagbaryon}) and (\ref{Lag:Q2A}). The results are
 \ba
  \Pi^{tt}_{\perp(1)} (p_0,\bar{p})
  &=&
  - \lambda^2 N_c
  [
    2 \bar{A}_0
    - (4 m_q^2 + \bar{p}^2 ) \bar{B}_0(p_0,\bar{p})\nonumber\\
&&    + \bar{B}^{tt}(p_0,\bar{p})
  ]
\ ,
\nonumber\\
  \Pi^{ts}_{\perp(1)} (p_0,\bar{p})
  &=&
  - \lambda^2 N_c
  \left[
    p_0 \bar{p} \, \bar{B}_0(p_0,\bar{p})
    + \bar{B}^{ts}(p_0,\bar{p})
  \right]
\ ,
\nonumber\\
  \Pi^{ss}_{\perp(1)} (p_0,\bar{p})
  &=&
   \lambda^2 N_c
  [
     2 \bar{A}_0
    - ( 4 m_q^2 - p_0^2 ) \bar{B}_0(p_0,\bar{p})\nonumber\\
   && - \bar{B}^{ss}(p_0,\bar{p})
  ]
\ , \nonumber\\
  \Pi^{tt}_{\perp(2)} (p_0,\bar{p})
  &=&
  -4 N_c \left( N_f c_{A1} + c_{A2} \right) m_q \bar{A}_0
\ ,
\nonumber\\
  \Pi^{ts}_{\perp(2)} (p_0,\bar{p})
  &=& 0
\ ,
\nonumber\\
  \Pi^{ss}_{\perp(2)} (p_0,\bar{p})
  &=&
  4 N_c \left( N_f c_{A1} + c_{A2} \right) m_q
  \bar{A}_0
\label{tensors}
\end{eqnarray}
where the subscript $(n)$ with $n=1,2$ represents the contribution
from the ${\cal O}(p^n)$ Lagrangian and
 \ba
  \bar{A}_0 \equiv
  - \int \frac{d^4k}{i(2\pi)^4} \Delta_D(k)
\ ,
\nonumber\\
  \bar{B}_0(p_0,\bar{p}) \equiv
  \int \frac{d^4k}{i(2\pi)^4}
  [
    \Delta_D(k) \Delta_F(k-p) \nonumber\\
    + \Delta_D(k-p) \Delta_F(k)
  ]
\ ,
\nonumber\\
  \bar{B}^{\mu\nu}(p_0,\bar{p}) \equiv
  \int \frac{d^4k}{i(2\pi)^4}
  (2k-p)^\mu (2k-p)^\nu\nonumber\\
  \left[
    \Delta_D(k) \Delta_F(k-p) + \Delta_D(k-p) \Delta_F(k)
  \right]
\ ,
\end{eqnarray}
Here $\Delta_D(k)$ and $\Delta_F(k)$ are given by
\begin{eqnarray}
  \Delta_D (k)
&\equiv&
  i \frac{\pi}{\omega_{m_q}(\bar{k})}
  \delta \left( k_0 - \omega_{m_q}(\bar{k}) \right)
  \theta \left( P_F - \bar{k} \right)
\ ,
\nonumber\\
  \Delta_F(k)
&\equiv&
  \frac{1}{ k^2 - m_q^2 + i \epsilon }
\ ,
\end{eqnarray}
with $P_F$ being the Fermi momentum of the quasiquark. Now an
explicit calculation gives
 \ba
- 2 \bar{A}_0 &=& \frac{1}{4\pi^2} \left[P_F \omega_F - m_q^2 \,
\ln \frac{ P_F + \omega_F }{m_q} \right] \ ,
\nonumber\\
\bar{B}_0(\bar{p},\bar{p}) &=& 0 \nonumber\ ,\\
\bar{B}^{tt}(\bar{p},\bar{p}) &=& \frac{1}{4\pi^2} \Biggl[
  - P_F \omega_F + m_q^2 \,\ln \frac{P_F+\omega_F}{m}\nonumber\\
  && + P_F^2 \, \ln \frac{\omega_F+P_F}{\omega_F-P_F}
\Biggr] \ ,
\nonumber\\
  \bar{B}^{ts}(p_0,\bar{p})
&=&
  \frac{p_0}{\bar{p}}
  \left[ \bar{B}_S - \bar{B}^{tt}(p_0,\bar{p}) \right]
\nonumber\ ,
\\
  \bar{B}^{ss}(p_0,\bar{p})
&=&
  - \left( 1 + \frac{p_0^2}{\bar{p}^2} \right) \bar{B}_S
  + \frac{p_0^2}{\bar{p}^2}\bar{B}^{tt}(p_0,\bar{p})
\nonumber\\
\bar{B}_S &\equiv& \frac{p_\mu p_\nu}{p^2}
\bar{B}^{\mu\nu}(p_0,\bar{p})=-2\bar{A}_0
\ .\label{explicit}
 \ea
Substituting (\ref{tensors}) with (\ref{explicit}) into
(\ref{fpit2: form}) and (\ref{fpits: form}), we find that there
are no hadronic dense loop contributions at ${\cal O}(p^4)$
from the ${\cal O}(p)$
Lagrangian (\ref{lagbaryon}):
\begin{eqnarray}
&&
  \delta_{(1)}\left[ f_\pi^t \right]^2
  =
  \overline{\Pi}^{tt}_{\perp(1)} (\bar{p},\bar{p})
  + \overline{\Pi}^{ts}_{\perp(1)} (\bar{p},\bar{p})
  = 0
\ ,
\nonumber\\
&&
  \delta_{(1)} \left[  f_\pi^t f_\pi^s \right]
  =
  - \overline{\Pi}^{ts}_{\perp(1)} (\bar{p},\bar{p})
  - \overline{\Pi}^{ss}_{\perp(1)} (\bar{p},\bar{p})
  = 0
\ .
\end{eqnarray}
As for contributions at ${\cal O}(p^5)$
from (\ref{Lag:Q2A}), the results are
\begin{eqnarray}
&&
  \delta_{(2)}\left[ f_\pi^t \right]^2
  =
  -4 N_c \left( N_f c_{A1} + c_{A2} \right) m_q \bar{A}_0
\ , \label{Fpt2:2A}
\\
&&
  \delta_{(2)}\left[  f_\pi^t f_\pi^s \right]
  =
  -4 N_c \left( N_f c_{A1} + c_{A2} \right) m_q \bar{A}_0
\ . \label{Fpts:2A}
\end{eqnarray}
In the small mass limit $m_q \ll P_F$, the corrections in
Eqs.~(\ref{Fpt2:2A}) and (\ref{Fpts:2A}) go as
\begin{eqnarray}
  \delta_{(2)}\left[ f_\pi^t \right]^2
  &=&
  \delta_{(2)}\left[  f_\pi^t f_\pi^s \right]\nonumber\\
  &&\mathop{\longrightarrow}_{m_q \ll P_F}
  N_c \left( N_f c_{A1} + c_{A2} \right) \,
  \frac{m_q \, P_F^2}{2\pi^2}
\ . \label{Fpi:2A:approx:2}
\end{eqnarray}
Thus at the fixed point $(g,a,m_q)=(0,1,0)$, the hadronic dense
loop corrections vanish. This means that we have no contributions
to $f_\pi^{t,s}$ from dense loop terms at the critical point.
Similarly there are no hadronic dense loop corrections to the pion
velocity, so we recover $v=1$ at the critical point.

\section{Vector Meson Mass}
\setcounter{equation}{0}
\renewcommand{\theequation}{\mbox{C.\arabic{equation}}}

In this appendix, we give details of the derivation of
Eq.(\ref{mrho at T 2}) that represents the hadronic
dense corrections to the vector meson mass. The relevant
piece of Lagrangian that is additional to (\ref{lagbaryon}) is of
${\cal O} (p^2)$ and has the form
\begin{eqnarray}
\delta {\cal L}_{Q(2V)} &=& \bar{\psi}\left\{
  c_{V1} \mbox{tr}
    \left[
      \hat{\alpha}_{\parallel\mu} \hat{\alpha}_{\parallel}^\mu
    \right]
  + c_{V2}
    \hat{\alpha}_{\parallel\mu}
    \hat{\alpha}_{\parallel}^\mu\right\}\psi\nonumber\\
  &&+\bar{\psi} c_{V3}
    \left[
      \hat{\alpha}_{\parallel\mu} \,,\, \hat{\alpha}_{\parallel\nu}
    \right]
    \left[ \gamma^\mu \,,\, \gamma^\nu \right]
\psi \ , \label{Lag:Q2V}
\end{eqnarray}
where $c_{V1}$, $c_{V2}$ and $c_{V3}$ are constants of mass
dimension $-1$~\footnote{For $N_f=2$, the $c_{V1}$-term and
$c_{V2}$-term are not independent. We can set, e.g., $c_{V2}=0$
without loss of generality.}.
The tadpole diagrams from the Lagrangian in Eq.~(\ref{Lag:Q2V}) generate
the ${\cal O}(p^5)$ corrections to the second term in
Eq.~(\ref{Lagrangian}).
The divergent corrections which are proportional to $m_q$ modify the
RGE for $F_\sigma$, and thus $a$.
At the critical density these corrections vanish since
$m_q \rightarrow 0$ for $\mu \rightarrow \mu_c$.

{}From the Lagrangian in Eqs.~(\ref{lagbaryon}) and
(\ref{Lag:Q2V}) the
vector-vector two-point function
$\Pi_V^{\mu\nu}$ gets the contributions
 \ba
\Pi_{V}^{\mu\nu\mbox{\scriptsize(tree)}}(p) &=& F_\sigma^2
g^{\mu\nu} - \frac{1}{g^2} \left( g^{\mu\nu} p^2 - p^\mu p^\nu
\right)\ , \label{treeV}\\
  \bar{\Pi}_{V(1)}^{\mu\nu}(p) &=&
  - (1-\kappa)^2
  \Bigl[
    \left( g^{\mu\nu}p^2 - p^\mu p^\nu \right)
    \bar{B}_0 (p_0,\bar{p})\nonumber\\
    &&+ 2 g^{\mu\nu} \bar{A}_0
     + \bar{B}^{\mu\nu} (p_0,\bar{p})
  \Bigr]
\ , \label{PiV:1}\\
 \Pi_{V(2)}^{\mu\nu}(p_0,\vec{p})
&=&
  2 N_c g^{\mu\nu} \left( N_f c_{V1} + c_{V2} \right)
  \left( - 2 \bar{A}_0 \right)
\ . \label{PiV:2}
\end{eqnarray}
As before, the subscript $(n)$ for $n=1,2$ represents the
contribution from the ${\cal O} (p^n)$ Lagrangian.

Let us define the vector meson mass through the general form
of the vector meson propagator at one-loop level.
In HLS at one-loop level,
$\Pi^{\mu\nu}_V (p_0,\vec{p})$,
which is related to the inverse propagator as in Eq.~(\ref{Dinv PiV}),
can be written as
\begin{eqnarray}
  \Pi^{\mu\nu}_V (p_0,\vec{p})
&=&
  F_\sigma^2 g^{\mu\nu} - \frac{1}{g^2}
  \left( g^{\mu\nu} p^2 - p^\mu p^\nu \right)
\nonumber\\
&&
  {}+ \overline{\Pi}^{\mu\nu}_V (p_0,\vec{p})
\ ,
\end{eqnarray}
where $\overline{\Pi}^{\mu\nu}_V (p_0,\vec{p})$ denotes the
hadronic dense/thermal corrections. Then the components in
Eq.~(\ref{decomp:PiV}) take the following form:
\begin{eqnarray}
&&
  \Pi_V^S(p_0;\bar{p})
  =
  F_\sigma^2
  + \bar{\Pi}_V^S(p_0;\bar{p})
\ ,
\nonumber\\
&&
  \Pi_V^L(p_0;\bar{p})
  =
  \frac{1}{g^2} p^2
  + \bar{\Pi}_V^L(p_0;\bar{p})
\ ,
\nonumber\\
&&
  \Pi_V^T(p_0;\bar{p})
  =
  \frac{1}{g^2} p^2
  + \bar{\Pi}_T^L(p_0;\bar{p})
\ ,
\nonumber\\
&&
  \Pi_V^C(p_0;\bar{p})
  =
  \bar{\Pi}_V^C(p_0;\bar{p})
\ .
\end{eqnarray}
Since $\Pi_V^C(p_0;\bar{p})$ part does not include ${\cal O}(1)$
contribution, we can neglect $\Pi_V^C(p_0;\bar{p})$ in the
denominator of the propagator in Eq.~(\ref{prop:form1}). Then the
propagator of the field $V_\mu$ is expressed as
\begin{eqnarray}
  - i D^{\mu\nu}
&=&
  P_L^{\mu\nu}
  \frac{1}{
    p^2/g^2 - F_\sigma^2
    + \left(
        \bar{\Pi}_V^L(p_0;\bar{p}) - \bar{\Pi}_V^S (p_0;\bar{p})
    \right)
  }
\nonumber\\
&&
  {}+
  P_T^{\mu\nu}
  \frac{1}{
    p^2/g^2 - F_\sigma^2
    + \left(
        \bar{\Pi}_V^T(p_0;\bar{p}) - \bar{\Pi}_V^S (p_0;\bar{p})
    \right)
  }
\nonumber\\
&&
  {}+
  P_C^{\mu\nu}
  \frac{\bar{\Pi}_V^C(p_0;\bar{p})}{
    F_\sigma^2 \left( p^2/g^2 - F_\sigma^2 \right)
  }\nonumber\\
&&
  {}+
  P_D^{\mu\nu}
  \frac{1}{ F_\sigma^2 + \bar{\Pi}_V^S (p_0;\bar{p}) }
\ . \label{prop:form2}
\end{eqnarray}
The vector-meson pole mass obtained from the pole of the
longitudinal propagator at its rest frame is
\begin{eqnarray}
&&m_{\rho L}^2 - M_\rho^2 =\nonumber\\
&-& g^2 \, \mbox{Re} \left(
  \bar{\Pi}_V^L(p_0=M_\rho;\bar{p}=0)
  - \bar{\Pi}_V^S (p_0=M_\rho;\bar{p}=0)
\right) \ , \nonumber\\
\end{eqnarray}
where $M_\rho = g F_\sigma$ is the tree level mass, and
$\mbox{Re}$ denotes the real part. Here $m_\rho$ is replaced by
$M_\rho$ in the loop corrections, since the difference is of
higher order. When one uses the transverse component, on the other
hand, the vector-meson pole mass is given by
\begin{eqnarray}
&&m_{\rho T}^2 - M_\rho^2=\nonumber\\
&& - g^2 \, \mbox{Re} \left(
  \bar{\Pi}_V^T(p_0=M_\rho;\bar{p}=0)
  - \bar{\Pi}_V^S (p_0=M_\rho;\bar{p}=0)
\right) \ .\nonumber\\
\end{eqnarray}

Consider now the ${\cal O}(p^4)$ correction summarized in
Eq.~(\ref{PiV:1}). Decomposing it into four components as in
Eq.~(\ref{decomp:PiV}), we get
\begin{eqnarray}
&&
  \bar{\Pi}_{V(1)}^S (p_0,\bar{p}) = 0
\ ,
\nonumber\\
&&
  \bar{\Pi}_{V(1)}^C (p_0,\bar{p}) = 0
\ ,
\nonumber\\
&&
  \bar{\Pi}_{V(1)}^L (p_0,\bar{p})
  =
  - (1-\kappa)^2
  \left[
    - p^2 \bar{B}_0(p_0,\bar{p}) + \bar{B}_L(p_0,\bar{p})
  \right]
\ ,
\nonumber\\
&&
  \bar{\Pi}_{V(1)}^T (p_0,\bar{p})
  =
  - (1-\kappa)^2
  \left[
    - p^2 \bar{B}_0(p_0,\bar{p}) + \bar{B}_T(p_0,\bar{p})
  \right]
\ .\nonumber\\
\end{eqnarray}
By using the formulas at rest
\begin{eqnarray}
  \bar{B}_T(p_0,\bar{p}=0)
  &=&
  \bar{B}_L(p_0,\bar{p}=0)\nonumber\\
  &=&
  \frac{2}{3}\, \bar{B}_S
  + \frac{ p_0^2 - 4 m_q^2}{3} \, \bar{B}_0(p_0,\bar{p}=0)
\  \label{BT:rest}
\end{eqnarray}
and
\begin{eqnarray}
&&
  \bar{B}_0(p_0,\bar{p}=0)
\nonumber\\
&& \quad
=
  \frac{1}{2} \int \frac{d^3\vec{k}}{(2\pi)^3}
  \frac{\theta(P_F - \bar{k})}{\omega_{m_q}(\bar{k})}
  \frac{1}{ p_0 ^2 - 4 \omega^2(\bar{k}) + i \epsilon }
\nonumber\\
&& \quad
  =
  \frac{1}{8\pi^2}
  \Biggl[
    - \ln \frac{ P_F + \omega_F }{ m_q }
    {}+ \frac{1}{2}
    \sqrt{ \frac{ 4m_q^2 -p_0^2 - i \epsilon }{ -p_0^2 - i \epsilon } }
\nonumber\\
&& \qquad\times
    \ln \frac{
      \omega_F\,\sqrt{ 4m_q^2 -p_0^2 - i \epsilon }
      + P_F\,\sqrt{ -p_0^2 - i \epsilon }
    }{
      \omega_F\,\sqrt{ 4m_q^2 -p_0^2 - i \epsilon }
      - P_F\,\sqrt{ -p_0^2 - i \epsilon }
    }
  \Biggr]
\ ,
\end{eqnarray}
 we obtain the corrections to the vector meson pole mass as
\begin{eqnarray}
&&
  \delta_{(1)} m_{\rho L}^2 = \delta_{(1)} m_{\rho T}^2
\nonumber\\
&& \quad
  = \frac{2}{3}\, g^2 (1-\kappa)^2
\nonumber\\
&& \qquad
\times
\Bigl[
  \bar{B}_S
  - (M_\rho^2 + 2 m_q^2) \,\mbox{Re}\,\bar{B}_0(p_0=M_\rho,0)
\Bigr]
\ .
\nonumber\\
\end{eqnarray}
When we take $M_\rho, m_q \ll P_F$ limit, the above expression
becomes
\begin{eqnarray}
\left. \delta_{(1)} m_{\rho L}^2 \right\vert_{M_\rho, m_q \ll P_F}
&=& \left. \delta_{(1)} m_{\rho T}^2 \right\vert_{M_\rho, m_q \ll
P_F}\nonumber\\
& =& \frac{g^2}{6\pi^2}\, (1-\kappa)^2 P_F^2 \ .
\end{eqnarray}

Next, let us include the higher order correction (${\cal O}(p^5)$).
{}From the corrections summarized in Eq.~(\ref{PiV:2}),
we obtain
\begin{eqnarray}
&&
  \bar{\Pi}_{V(2)}^S (p_0,\bar{p}) =
  2 N_c \left( N_f c_{V1} + c_{V2} \right)
  \bar{B}_S
\ ,
\nonumber\\
&&
  \bar{\Pi}_{V(2)}^C (p_0,\bar{p}) =
  \bar{\Pi}_{V(2)}^L (p_0,\bar{p}) =
  \bar{\Pi}_{V(2)}^T (p_0,\bar{p}) =
  0
\ .
\end{eqnarray}
Then, the corrections to the vector meson pole masses are
\begin{eqnarray}
\delta_{(2)} m_{\rho L}^2 = \delta_{(2)} m_{\rho T}^2 = g^2 \, 2
N_c \left( N_f c_{V1} + c_{V2} \right) \bar{B}_S \ .
\end{eqnarray}
In the $M_\rho, m_q \ll P_F$ limit, this expression is reduced to
\begin{eqnarray}
&&
  \left. \delta_{(2)} m_{\rho L}^2 \right\vert_{M_\rho, m_q \ll P_F}
  =
  \left. \delta_{(2)} m_{\rho T}^2 \right\vert_{M_\rho, m_q \ll P_F}
\nonumber\\
&& \qquad\quad
  = \frac{g^2}{2\pi^2} \, N_c \left( N_f c_{V1} + c_{V2} \right)
\, P_F^2 \ .
\end{eqnarray}
Note that up to ${\cal O}(p^6)$ corrections,
the longitudinal and transverse
pole masses are the same. This is the reason for the Lorentz
invariant structure of the $\rho$ mass in Eq.(\ref{mrho at T 2}).

\section{MK Theorem}
\setcounter{equation}{0}
\renewcommand{\theequation}{\mbox{D.\arabic{equation}}}

In this appendix we sketch how to go from the RGEs in $\M$ to the
RGEs in $\mu$~\cite{morley-kislinger}.  As noted in the text, the
reasoning is applicable to fundamental theories such as QED or QCD
(in weak-coupling sector), but not without modifications to
effective theories such as HLS except perhaps for low temperature
or low density.

Denote the renormalized thermodynamic potential $\Omega_R
(h_i^R,F_\pi^R, m_q^R,\mu,\M)$ where $h_i$ stands generically for
$h_1=g$, $h_2=a$ and $h_3=\lambda$ (that is, $\Omega (h_i)$ stands
for $\Omega (h_1, h_2, h_3)$) , $\mu$ is the chemical potential
and $\M$ is the renormalization scale parameter. The RG invariance
condition that $\M \frac{d}{d\M} \Omega_R=0$ gives
 \ba
&& \Biggl[
  \M\frac{\partial}{\partial \M}
  +\beta (h_i^R)  \frac{\partial}{\partial h_i^R}
  - m_q^R\gamma_{m} \frac{\partial}{\partial m_q^R}
  - F_\pi^R\gamma_f \frac{\partial}{\partial F_\pi^R}
\Biggr] \no && \qquad \times\Omega_R (h_i^R,F_\pi^R,m_q^R,\mu,\M)
=0 \ , \label{aha-1}
 \ea
where
 \ba
\beta (h_i^R)&=& \M\frac{\partial h_i^R}{\partial \M} \ ,\no
\gamma_m&=&-\frac{1}{m_q^R}\M\frac{\partial m_q^R}{\partial
\M},\no \gamma_f&=&-\frac{1}{F^R_\pi}\M\frac{\partial
F_\pi}{\partial \M}
 \ea
Since  the thermodynamic potential (or pressure) has a mass
dimension 4, it should satisfy the identity
 \ba
&& \Biggl[
  \M\frac{\partial}{\partial \M}
  + \mu\frac{\partial}{\partial \mu}
  + m_q^R \frac{\partial}{\partial m_q^R}
  +F_\pi^R\frac{\partial}{\partial F_\pi^R}
\Biggr] \no && \qquad \times\Omega_R (h_i^R, F_\pi^R,m_q^R,\mu,\M)
\nonumber\\
&& \quad
 =4\Omega_R (h_i^R,
F_\pi^R,m_q^R,\mu,\M) \ . \label{id1-1} \ea By combining
(\ref{aha-1}) and (\ref{id1-1}), we can obtain \ba
[\mu\frac{\partial}{\partial \mu} -\sum_i\beta (h_i^R)
\frac{\partial}{\partial h_i^R}+m_q^R(1+\gamma_{m_q^R})
\frac{\partial}{\partial m_q^R} \no +F_\pi^R(1+\gamma_f)
\frac{\partial}{\partial F_\pi^R} -4] \Omega_R (h_i^R,
F_\pi^R,m_q^R,\mu,\M) =0.\label{rge2-1} \ea In the low density
region we expect that the ``intrinsic'' density dependence of the
bare theory is small, and thus we introduce the following ansatz
\ba && \Omega_R (h_i^R, F_\pi^R,m_q^R,\mu,\M)
\nonumber\\
&& \qquad =\mu^4\bar\Omega_R
 (h_i^R,
F_\pi^R,m_q^R,\mu,\M)\label{sol1-1}.
 \ea
Then $\bar\Omega_R$ satisfies
 \ba
&& \Biggl[
  \mu\frac{\partial}{\partial \mu}
  -\sum_i\beta (h_i(\mu)) \frac{\partial}{\partial h_i (\mu)}
  + m_q(\mu)(1+\gamma_{m}) \frac{\partial}{\partial m_q (\mu)}
\no && \qquad
 +F_\pi (\mu) (1+\gamma_f(\mu) ) \frac{\partial}{\partial F_\pi (\mu)}
\Biggr] \no && \quad \times \bar\Omega_R ( h_i(\mu), F_\pi (\mu),
m_q(\mu),\mu, \M) =0 \ , \label{sol2-1}
 \ea
where,
 \ba \mu\frac{\partial h_i(\mu)}{\partial\mu} &=&\beta_i
(h_i(\mu))\no \mu\frac{\partial m_q (\mu)}{\partial\mu}
&=&-[1+\gamma_m (h_i(\mu))]m_q(\mu)\no \mu\frac{\partial F_\pi
(\mu)}{\partial\mu} &=&-[1+\gamma_f (h_i(\mu))]F_\pi
(\mu)\label{hhhg}
 \ea
with the conditions
 \ba
h_i (\mu)|_{\mu=\M} = h_i^R
 \ea
etc. From (\ref{sol2-1}) and (\ref{hhhg}),
 Eqs.(\ref{RGE-mu}) follow. For instance
we have for $F_\pi$
 \ba
\mu\frac{\partial F_\pi^2}{\partial
\mu}&=&-2F_\pi^2[1-\frac{1}{2F_\pi^2}\mu \frac{\partial
F_\pi^2}{\partial \mu}]\no &=& -2F_\pi^2 +C[3a^2g^2F_\pi^2
+2(2-a)\mu^2] \no &-& \frac{m_q^2}{2\pi^2}\lambda^2 N_c.
 \ea

\end{document}